\documentclass[pre,showpacs,twocolumn,groupedaddress,superscriptaddresst]{revtex4-1}

\usepackage{graphicx,amssymb,color}
\usepackage[dvipsnames]{xcolor}

\newcommand{\beq}{\begin{equation}}
\newcommand{\eeq}{\end{equation}}
\newcommand{\beqs} {\begin{displaymath}}
\newcommand{\eeqs} {\end{displaymath}}
\newcommand{\beqa} {\begin{eqnarray}}
\newcommand{\eeqa} {\end{eqnarray}}
\newcommand{\beqas} {\begin{eqnarray*}}
\newcommand{\eeqas} {\end{eqnarray*}}
\newcommand{\reff}[1]{(\ref{#1})}
\newcommand{\Vec}[1]{{\bf #1}}
\newcommand{\derpar}[2]{\frac{\partial #1}{\partial #2}}

\begin{document}
\title{Dynamics of a first order transition to an absorbing state}

\author{Baptiste N{\'e}el}
\altaffiliation[Present address: ] {D\'epartement de Physique, Ecole Normale Sup\'erieure, 24 rue Lhomond
75005 Paris, France}

\author{Ignacio Rondini}

\author{Alex Turzillo}
\altaffiliation[Present address: ] {California Institute of Technology, 1200 East California Boulevard
Pasadena, CA 91125, USA}

\author{Nicol\'as Mujica}

\author{Rodrigo Soto}
\email[]{rsoto@dfi.uchile.cl}
\affiliation{Departamento de F\'{\i}sica, Facultad de Ciencias F\'{\i}sicas y Matem\'aticas Universidad de Chile,
Avenida Blanco Encalada 2008, Santiago, Chile}

\date{\today}

\begin{abstract}
A granular system confined in a quasi two-dimensional box that is vertically vibrated can transit to an absorbing state in which all particles bounce vertically in phase with the box, with no horizontal motion. In principle, this state can be reached for any density lower than the one corresponding to one complete monolayer, which is then the critical density. Below this critical value, the transition to the absorbing state is of first order, with long metastable periods, followed by rapid transitions driven by homogeneous nucleation. Molecular dynamics simulations and experiments show that there is a dramatic increase on the metastable times far below the critical density; in practice, it is impossible to observe spontaneous transitions close to the critical density. This peculiar feature is a consequence of the non-equilibrium nature of this first order transition to the absorbing state. A Ginzburg-Landau model, with multiplicative noise, describes qualitatively the observed phenomena and explains the macroscopic size of the critical nuclei. The nuclei become of small size only close to a second critical point where the active phase becomes unstable  via a saddle node bifurcation. It is only close to this second critical point that experiments and simulations can evidence spontaneous transitions to the absorbing state while the metastable times grow dramatically moving away from~it.

\end{abstract}
\maketitle

\section{Introduction}

Granular matter has been considered for some time as a prototype of non-equilibrium systems, which serves as playground for generic non-equilibrium dynamics. The need of permanent energy injection into granular matter to sustain dynamic states places these systems under intrinsic non-equilibrium conditions. Energy is dissipated at grain-grain and grain-boundary collisions but it is also injected by different mechanisms, typically via boundaries or external fields acting on the bulk and, therefore, the detailed balance condition is not fulfilled.

The dissipative and macroscopic nature of granular matter creates difficulties in their detailed study and comparison with other non-equilibrium systems. Firstly, energy is normally injected non-homogeneously, through boundaries and large inhomogeneities usually develop. For example, shear bands of only a few grain diameters in size spontaneously develop or closed packed clusters form, coexisting with dilute regions. Secondly, it is very difficult to observe experimentally the interior of the flow and only surface information is available. Only in some cases bulk dynamics has been studied using sophisticated experimental methods with limited information \cite{NMR,PEPT,IndexMatch}.

Recently, the quasi two dimensional (Q2D) geometry, in which grains are placed in a box of height smaller that two grain diameters and large horizontal dimensions, has gained attention as it overcomes these two difficulties. When the box is vertically vibrated, energy is transferred to the vertical degree of freedom of the grains as they collide with the top and bottom walls. Subsequent collisions among grains transfer this energy to the horizontal degrees of freedom. The system, when observed from above and considering only the two-dimensional dynamics, resembles a fluid. Here, we will focus on this effective two-dimensional system. This dimensional reduction is appropriate as the dynamical time scale of interest is larger than the fast vertical motion, which acts as an energy source \cite{BritoRissoSoto}. 
In the Q2D geometry, there is a large range of parameters for which the system remains in stable homogeneous states \cite{olafsen,prevost2004,Melby2005}, both in dense or dilute phases. Also, in experiments it is possible to follow in detail the motion of all grains, allowing for a simultaneous analysis of the collective and individual dynamics. 

Varying the filling fraction and the box vibration amplitude and frequency, the system develops a solid-liquid transition, with different possible solid phases \cite{olafsen,prevost2004,Melby2005}. This phase  separation is triggered by the negative compressibility of the effective 2D equation of state and the associated transient dynamics is governed by waves \cite{Argentina,Clerc}. Also, the analysis of an orientational order parameter has shown that in different regions of the phase space the transition can be either of first or second order type. In the second order case, critical phenomena has been reported \cite{Castillo}. Other phase transitions have also been observed in this system \cite{Superheating,explosion}.
Finally, for a similar Q2D geometry but with no top lid, an analysis based on fluctuating hydrodynamics has been performed and the fluctuation theorems have been analyzed in detail~\cite{puglisi1,puglisi2}.

In this article, we concentrate in a non-equilibrium phase transition to an absorbing state in the Q2D geometry. Absorbing states are particular non-equilibrium states characterized by total freezing of the relevant degrees of freedom, where fluctuations are absent completely \cite{AbsReview1,AbsReview2,AbsReview3,AbsReview4}. Then, if the system transits to an absorbing state it will remain in this state indefinitely as there are no mechanisms that can take it out from it. There are numerous examples of absorbing states in nature including population dynamics \cite{Albano}, epidemics \cite{Mollison}, ecological models \cite{Satulovsky}, turbulent liquid crystals \cite{Takeuchi} and surface catalytic reactions~\cite{Ziff}. 
The absorbing transition presents a critical point that has deserved large attention. It shows interesting critical properties and has  become a prototype for non-equilibrium phase transitions, allowing the construction of non-equilibrium universality classes.
The absorbing transition can also be of first order (discontinuous), as shown in \cite{Ziff}, and the dynamical properties of this case has been studied in one dimension in Ref. \cite{Meerson}. 
In this article we show that the non-equilibrium character of the absorbing transition gives rise to interesting properties, even in the first order case as the dramatic increase of the critical nuclei close to the transition \cite{Meerson}.

The transition presented in this article is between a fluidized state with grains having horizontal motion and a state with particles moving vertically in phase with the box. One important feature is that spontaneous transitions only occur deep into the coexistence region and, in practice, it is impossible to observe them close to the critical point. Only by means of initially seeded nuclei it is possible to study the system's properties near the critical point. This feature is a direct consequence of a transition to a non-equilibrium absorbing state, as it is shown using a Ginzburg-Landau model \cite{Meerson}.

The article is organized as follows. Section \ref{sec.Q2D} describes the Q2D geometry and its main properties, specially, the existence of a fixed point absorbing state. In~Sec. \ref{sec.MDresults} the main results concerning the phase transition to the absorbing state obtained using molecular dynamics simulations are presented. The transition is characterized starting from both homogeneous states or with  preformed nuclei. Experimental results of the same geometry, but with a different control parameter are presented in Sec.~\ref{sec.experiments}. Based on the results, a continuous Ginzburg-Landau-like model is built and compared with the simulations and experiments in Sec.~\ref{sec.landau}. Finally, conclusions and perspectives are given in Sec.~\ref{sec.conclusions}.
 
\section{Quasi 2D fluidized granular matter} \label{sec.Q2D}

The system under consideration consists on $N$ spherical grains of mass $m$ and diameter $\sigma$ placed in a shallow box of size $L_x\times L_y\times L_z$, with the height less than two particle diameters, while the lateral directions are large. This Q2D box is vertically vibrated with a fixed amplitude $A$ and angular frequency $\omega$. The vibration injects energy to the vertical motion of the grains that, through grain-grain collisions, transfer this energy to the horizontal motion. Grain-grain and grain-wall collisions are dissipative and, therefore, the system can achieve non-equilibrium steady states with a finite average kinetic energy. 
In these fluid states, the vertical and in-plane motion are very different and energy equipartition is not satisfied: the vertical kinetic energy is typically larger than the horizontal one~\cite{anisotropia1,anisotropia2}. 
In what follows we will consider only the effect of the top and bottom walls and disregard the effect of the lateral walls. In simulations this is achieved by using periodic boundary conditions in the lateral directions, while in experiments large boxes will be used such that the effect of the lateral walls is limited to a finite boundary layer.

If the grain-wall collision has friction, a single grain put in the box can rapidly achieve a state in which it has a finite vertical energy while there is no horizontal motion or rotation. Seen from above, the grain is in a fixed position. Depending on the amplitude, frequency and the collision parameters, the vertical motion can be periodic or chaotic. In the region of parameters we will consider, which are experimentally feasible, the motion is periodic moving in phase with the box with the same period. We call this state the {\em fixed point} which can have important dynamical effects \cite{explosion,explosion2}. 

Being the fixed point unique, if several grains are placed far apart in the box, they will eventually reach the fixed motion and move synchronously. When all the grains are in the fixed point, the horizontal kinetic energy vanishes and therefore grains cannot collide among them. There are no fluctuations either and the grain motion is purely vertical. Therefore, once the system reaches this state with all grains in the fixed point it cannot go back to a fluidized regime with horizontal energy. It is then an example of an {\em absorbing state}. 

As indicated, it is expected that at low particle densities, the system can reach the absorbing state as grain-grain collisions are inefficient compared to the attraction to the fixed point. On the other hand, if the number of particles is large enough, such that a single monolayer of grains can not be accommodated,  the absorbing state can not be reached. In the case of an infinite system, the maximum number density is $n_{\max}\equiv N\sigma^2/(L_x L_y) = 2/\sqrt{3}$ for an hexagonal close packing. 
At lower densities, but still under dense conditions, it is possible prepare special initial conditions with all grains sufficiently close to their fixed points. It is clear that the the system will be attracted to  the absorbing state. The attraction basin is, however, small and unless the system is specially prepared, it will be extremely rare to fall to the absorbing state. In this article we study the dynamics of the transition to the absorbing state and relate this dynamics with the shrinking of the attraction basin. 

\section{Molecular dynamics simulations} \label{sec.MDresults}
\subsection{Model parameters}
The Q2D system described in the previous section is studied first by means of molecular dynamics simulations using the event driven algorithm \cite{ED}. Collisions are instantaneous and the inelasticity is taken into account using the inelastic hard sphere model (IHS): there are normal and tangential restitution coefficients ($\alpha_n$ and $\alpha_t$) and static and dynamic friction coefficients ($\mu_s$ and $\mu_d$). All these coefficients are considered constants, independent on the colliding velocity, and the same for the grain-grain and grain-wall collisions. Specifically, we use $\alpha_n=\alpha_t=0.87$, $\mu_s=0.16$, and $\mu_d=0.13$, which give good quantitative agreement with experiments \cite{Clerc}. Other values give similar results to the ones reported here. The box lateral dimensions are $L_x=L_y=100\sigma$ while the height is $L_x=1.8\sigma$. The whole box is vibrated with a bi-parabolic waveform at amplitude $A=0.064\sigma$ and frequency $\omega=8.56 \sqrt{g/\sigma}$, where $g$ is the acceleration of gravity \cite{biparabolic}.  
The two microscopic time scales of the system, the one given by gravity $\sqrt{\sigma/g}$ and the oscillation period $T=2\pi/\omega$, are of the same order ($T\approx 0.73 \sqrt{\sigma/g}$). Therefore, the microscopic dynamic of the system is characterized by a unique time scale that we chose to be the oscillation period. Initial conditions, unless otherwise stated, are created with grains homogeneously placed in the box, with random linear and angular velocities. Details of the initial condition are rapidly lost and the system reaches an homogeneous fluidized state. 

\begin{figure}[t!]
\includegraphics[width=.9\columnwidth]{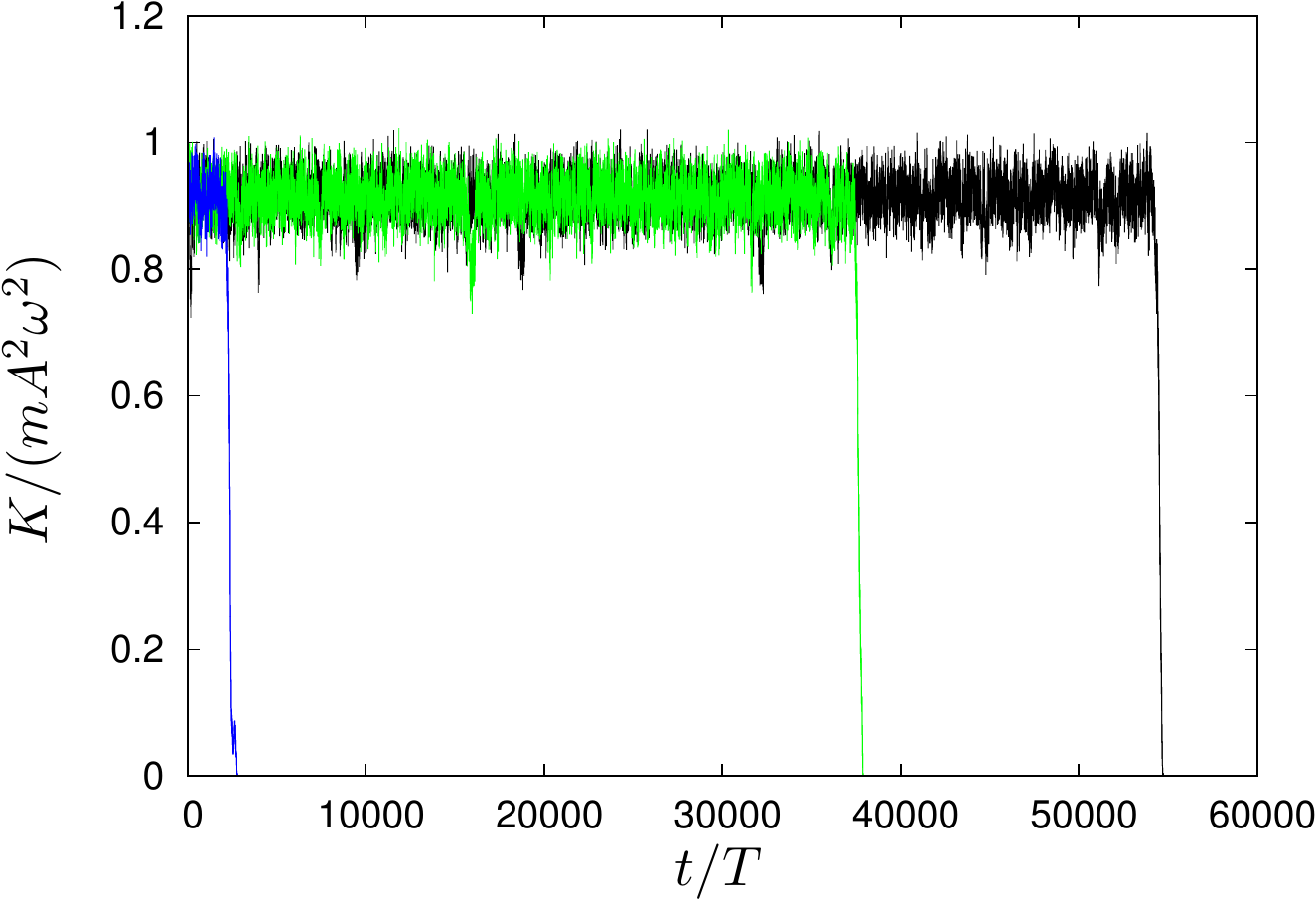}
\caption{(color online)  Normalized horizontal kinetic energy per particle as a function of time. The three simulations shown in black, blue (dark gray) and green (light gray) have the same number of particles, $N=5160$, and differ only on grain's initial random conditions.}
\label{fig.energy}
\end{figure}

\subsection{Homogeneous nucleation}
A series of simulations is done with different number of particles. For each value of $N$, several simulations are performed, changing randomly the initial positions and velocities of the particles, while keeping spacial homogeneity. 
Fig.~\ref{fig.energy} shows the time evolution of the normalized horizontal kinetic energy per particle $K$ for various simulations done with $N=5160$. After a short transient, the kinetic energy rapidly reaches a steady steady value. Then, after a transition time $\tau$, that varies from simulation to simulation, the kinetic energy rapidly drops and the system transits to the absorbing state, with vanishing horizontal energy. 

\begin{figure}[t!]
\includegraphics[width=.98\columnwidth]{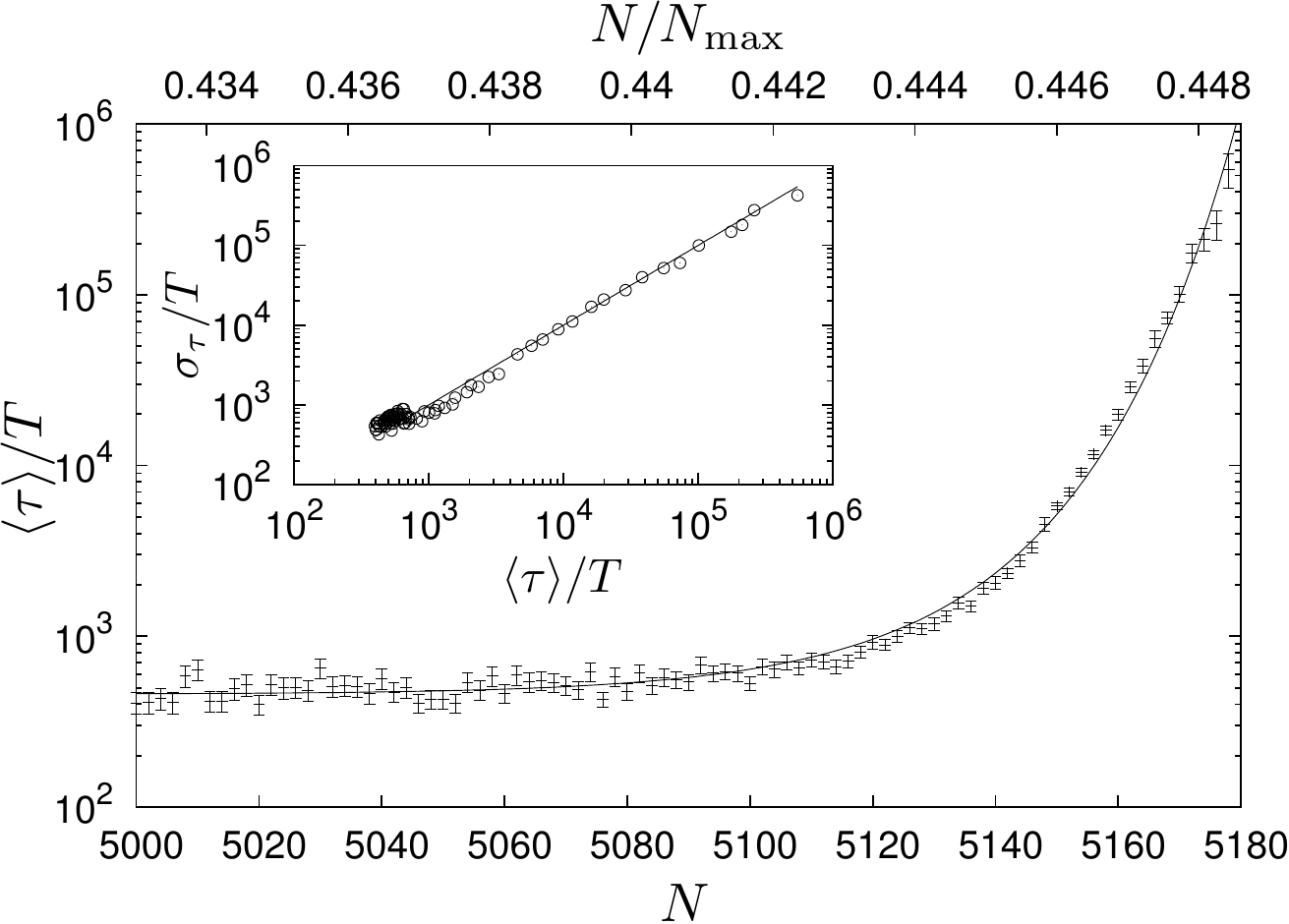}
\caption{Average transition time $\langle \tau \rangle$ to the absorbing state as a function of the number of particles $N$ (normalized to the $N_{\rm max}$ in the top axis). The average and errorbars are computed from $N_r$ independent realizations. For the later we use the standard error definition $\sigma_\tau/\sqrt{N_r}$, where $\sigma_\tau$ is the standard deviation of the $N_r$ measurements of the fluidization time $\tau$.  The solid line corresponds to the double exponential fit \reff{fit.sims}.
The inset shows the standard deviation $\sigma_\tau$ versus $\langle\tau\rangle$. The straight line represents the prediction for a Poissonian process, $\sigma_\tau=\langle\tau\rangle$.}
\label{fig.tauvsN} 
\end{figure}

\begin{figure*}
\includegraphics[width=1.98\columnwidth]{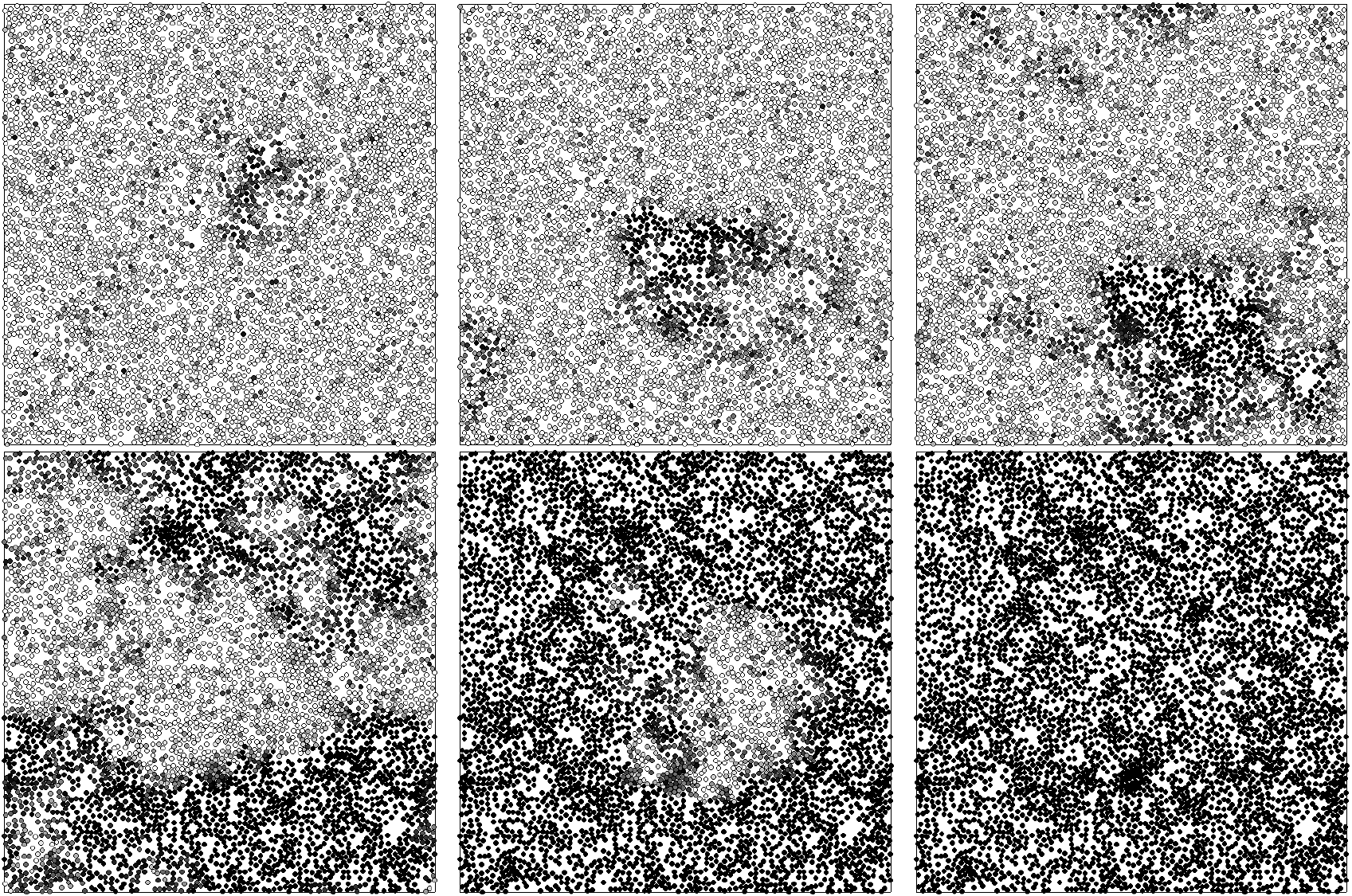}
\caption{Sequence of snapshots of the system as it nucleates to the absorbing state. The gray intensity is proportional to the horizontal kinetic energy of each grain, with black grains having vanishing horizontal energy. The time difference between snapshots is $\Delta t = 73.4 T$. The number of particles is $N=5128$. For this number of particles, the average transition time is $\langle\tau\rangle\approx1151 T$. 
The full video can be found in the Supplemental Material \cite{supmat}.}
\label{fig.Nucleation}
\end{figure*}

The transition times show a large variability with an average $\langle\tau\rangle$ shown in Fig. \ref{fig.tauvsN} that presents a rapid increase at $N\sim 5130$. However, there appears not to be a divergence at a finite number of particles but rather a rapid increase that is well fitted by a double exponential  expression 
\beq
\langle \tau \rangle/T = \exp\left[ a+\exp[{ c (N- N_0)]} \right], \label{fit.sims}
\eeq
with $ a = 6.13\pm 0.02$, $ c=0.0394\pm 0.0007$, and $ N_0= 5127 \pm 1$.  
We recall that the explored region of $N$ is far below the maximum number of particles that can be simultaneously in the fixed point, namely one monolayer in hexagonal close packing $N_{\rm max}=11547$.
For $N<N_0$ the transition time is almost constant. There is no metastability and the transition time is purely associated to the decay time to the absorbing state.

The variability in the transition times is computed by means of its standard deviation $\sigma_\tau=\sqrt{\langle\tau^2\rangle - \langle\tau\rangle^2}$, which is plotted against the average transition time in  Fig.~\ref{fig.tauvsN} inset. For large transition times, $\sigma_\tau = \langle \tau\rangle$, characteristic of a Poissonian processes. Higher moments of the distribution of times are difficult to obtain to show unanbigously that the process is Poissonian.

First order transitions are normally triggered by a nucleus in the final phase (the absorbing state). Indeed, direct observation of the grain motion in the simulation confirms that the transition is driven by a nucleation process. A spontaneous fluctuation creates a region of particles with small horizontal kinetic energy, almost in the fixed point. If the nucleus is large enough, it grows until the whole system reaches the fixed point. Small nuclei decay rapidly and the system goes back to the fluid (metastable) state. A sequence of the nucleation processes is shown in Fig.~\ref{fig.Nucleation} and a detailed video is available~\cite{supmat}.  Critical nuclei are not circular, neither the subsequent growing domains. This is an effect of large fluctuations and small effective surface tension between the fluid phase and the absorbing phase~\cite{Luu2012}.

Our results show that the transition is from a metastable state to an absorbing state. The transition process is Poissonian, which is triggered by a short memory nucleation process. The rapidly growing transition times are consistent with increasingly smaller probabilities of spontaneously forming the critical nucleus. 

\subsection{Initial nucleus}
The rapid increase of the transition times at a relatively small value of $N$ compared to $N_{\rm max}$ is intriguing. In practice, no simulation with $N\geq5180$ reached the absorbing state as long as the simulation lasted ($t_{\rm max}=10^6 T$). However, we expect that only for $N>N_{\rm max}$ the transition to the absorbing state is forbidden.  

To clarify this point, simulations with an initial nucleus were run until they either reach the absorbing state or go to a fluidized state. For each number of particles, an initial nucleus of radius $R$ is placed at a random position, inside which all particles are placed in the fixed point. Namely, their vertical position and velocities are those of the fixed point, while their horizontal and angular velocities vanish. For each $N$ and $R$,  several simulations varying the nucleus initial position are run for a maximum time $t_{\rm nucl}=1400 T$. The fraction $\lambda$ of simulations that reach the absorbing state is recorded.

\begin{figure}[t!]
\includegraphics[width=.85\columnwidth]{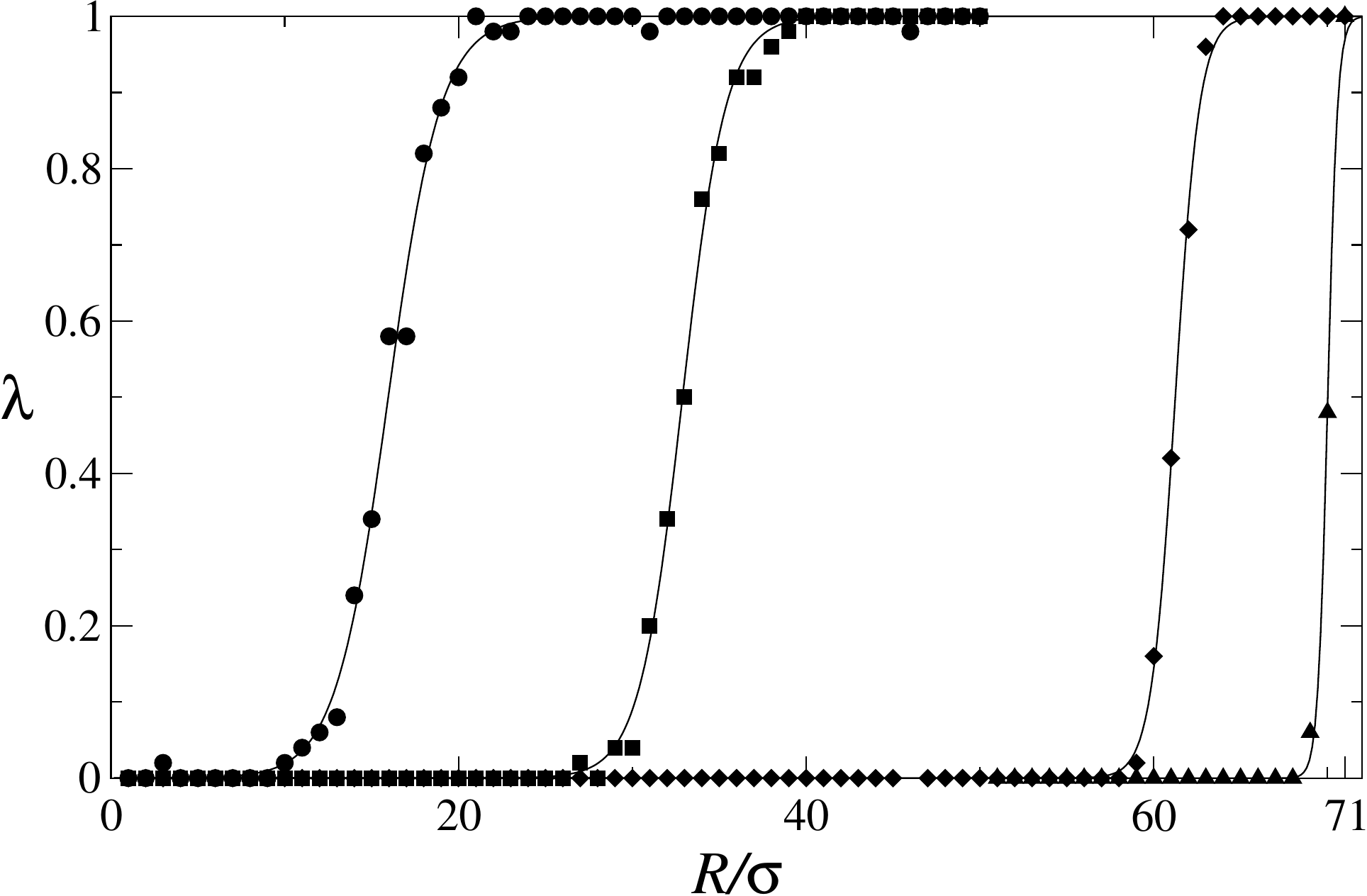}
\caption{Transition probability to the absorbing state starting with a nucleus seed of radius $R$. Symbols correspond to the results of the molecular dynamics simulations for different number of particles in the box. From left to right $N=5170$, $5300$, $6000$, and $8000$. Continuous lines are the result of fits  for each case using expression (\ref{eq.fitlambdaR}). }
\label{fig.transitionprob}
\end{figure}

Fig. \ref{fig.transitionprob} shows $\lambda$ as a function of the nucleus radius for various number of particles. 
There is an abrupt transition separating the cases of small radii for which the systems always re-fluidize from the cases of nuclei large enough to attract the whole system to the absorbing state. Note that for a box size of size $L_x\times\ L_y=100\times100\sigma^2$, the maximum radius size is $R_{\rm max}=50\sqrt{2}\sigma\approx 71 \sigma $, in which case all the particles would be already in the fixed point and $\lambda=1$. 
 The critical radius $R_c$  for each number of particles is obtained fitting
\beq
\lambda(R)=\frac{1}{2}\left[\tanh\left(\frac{R-R_c}{\Delta} \right) +1\right], \label{eq.fitlambdaR}
\eeq
where $\Delta$ is identified as a transition width. Figure \ref{fig.criticalradius} shows the critical radius as a function of the number of particles in the box. For $N \lesssim 5100$, compatible with the value of $N_0$ in \reff{fit.sims},  the system reaches the absorbing state without needing an initial nucleus, therefore the critical radius vanishes. Then, the critical radius rapidly increases to values close to the maximum radius given the box size. It is remarkable that already at $N=6000$, far below $N_{\rm max}$, the critical nucleus occupies $90\%$ of the total area.

\section{Experimental characterization} 
\label{sec.experiments}

\subsection{Experimental setup and protocols}

\begin{figure}[t!]
\includegraphics[width=.98\columnwidth]{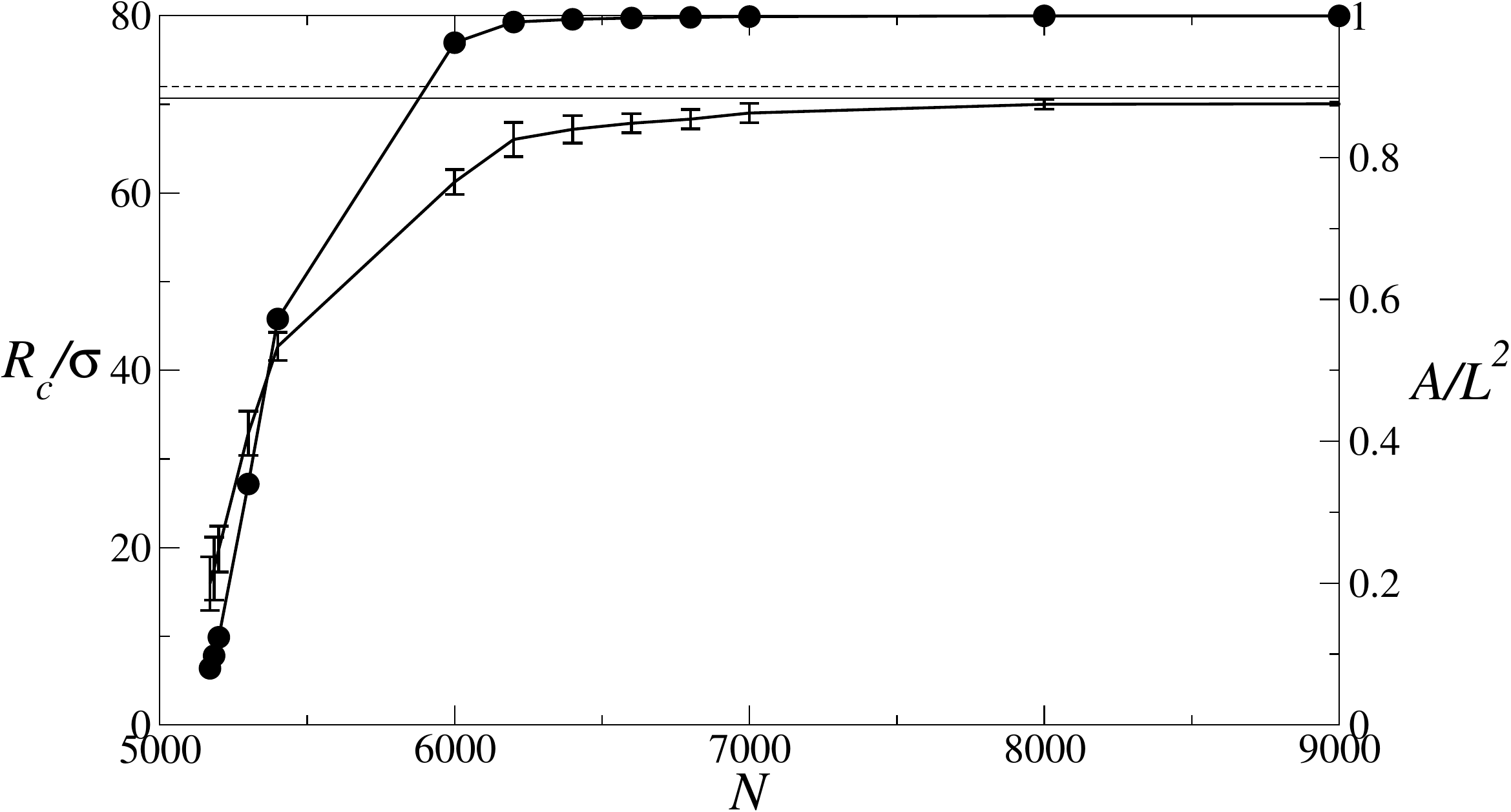}
\caption{
Radius (error bars, left axis) and relative area (solid circles, right axis) of the critical nuclei as a function of the number of particles. The error bars correspond to the width $\Delta$ of the transition probability fit according to Eq. (\ref{eq.fitlambdaR}). The connecting solid lines are placed as a guide. The horizontal solid line indicates the maximum radius allowed by the box size, $R_{\rm max}=50\sqrt{2}\sigma$, while the horizontal dashed line corresponds to $90\%$ of the total box area.}
\label{fig.criticalradius}
\end{figure}

\begin{figure*}[t!]
\begin{center}
\includegraphics[width=1.98\columnwidth]{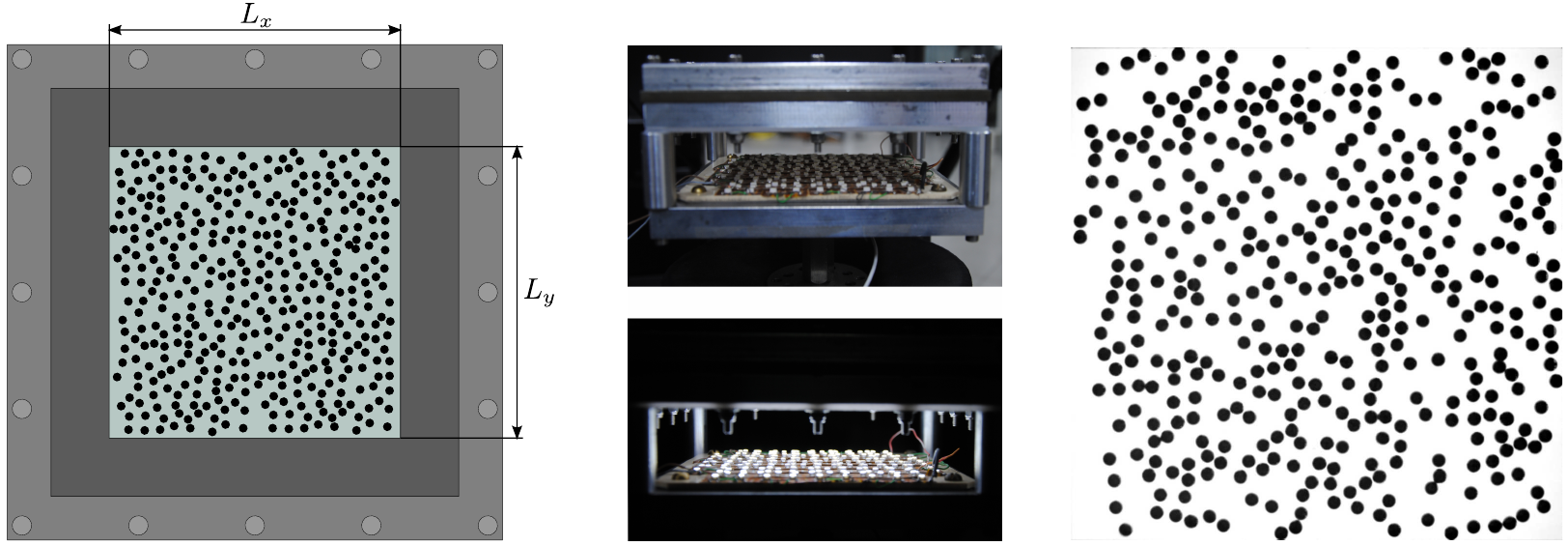}
\caption{Details of the experimental setup. (a) Schematic illustration of the experimental cell, showing the horizontal dimensions and the frame that holds the bottom and top glass plates; (b) Side view pictures of the cell and the base with the illumination LED array off (above) and on (below); (c) Typical image used for particle position detection. From two consecutive images particle velocities are measured as well.}
\label{Exp_Cell}
\end{center}
\end{figure*}

Experiments in the quasi 2D geometry have been performed to identify and characterize the transition to the absorbing state. Details of the experimental cell are shown in Fig. \ref{Exp_Cell}. The experimental setup and procedures are similar to the ones we have used previously \cite{Castillo,explosion}. We consider $N$ stainless steel spherical particles of diameter $\sigma = 3$~mm placed in a box of dimensions $L_x \times L_y \times L_z = 33.3 \times 33.3 \times 1.8 \  \sigma^3$. 
The horizontal dimensions are limited by the dimensions of the external frame, the internal frame being as large as possible to study an assembly of grains as large as possible (Fig. \ref{Exp_Cell}a). The upper limit of $N$ is imposed by the solid-liquid-like phase transition observed at high density~\cite{Castillo}. Based on preliminary numerical results and after some adaptation to the experimental case, we fix the beads filling fraction to $\phi = N \pi \sigma^2/(4 L_x L_y) = 0.28$, with $N = 400$.

The experimental cell is vertically vibrated by means of an electro-mechanical shaker, with displacement $z(t) = A \sin(\omega t)$. The driving frequency is $f = \omega/2\pi= 1/T = 70$ Hz. The vibration amplitude is measured with a piezoelectric accelerometer mounted to the cell's base. The dimensionless acceleration $\Gamma = A\omega^2/g$ is varied in the range $1-3$. The bottom and top walls are 10 mm thick glass plates. As collisions between particles and the glass walls can create electrical charges on the surfaces, the inner surface of both glass plates are coated by a thin indium tin oxide (ITO) film, which dissipates charges minimizing electrostatic effects. An array of light emitting diodes (LED) is placed below, fixed on the cell's base, and a white acrylic sheet is used in order to diffuse light (Fig. \ref{Exp_Cell}b). Thus, when viewed from above, particles are seen as black disks over a white background (Fig. \ref{Exp_Cell}c). Top view images are obtained with a high speed camera. Particle positions are determined at subpixel accuracy. From two consecutive images particle velocities are measured as well.

\begin{figure}[t!]
\includegraphics[width=.9\columnwidth]{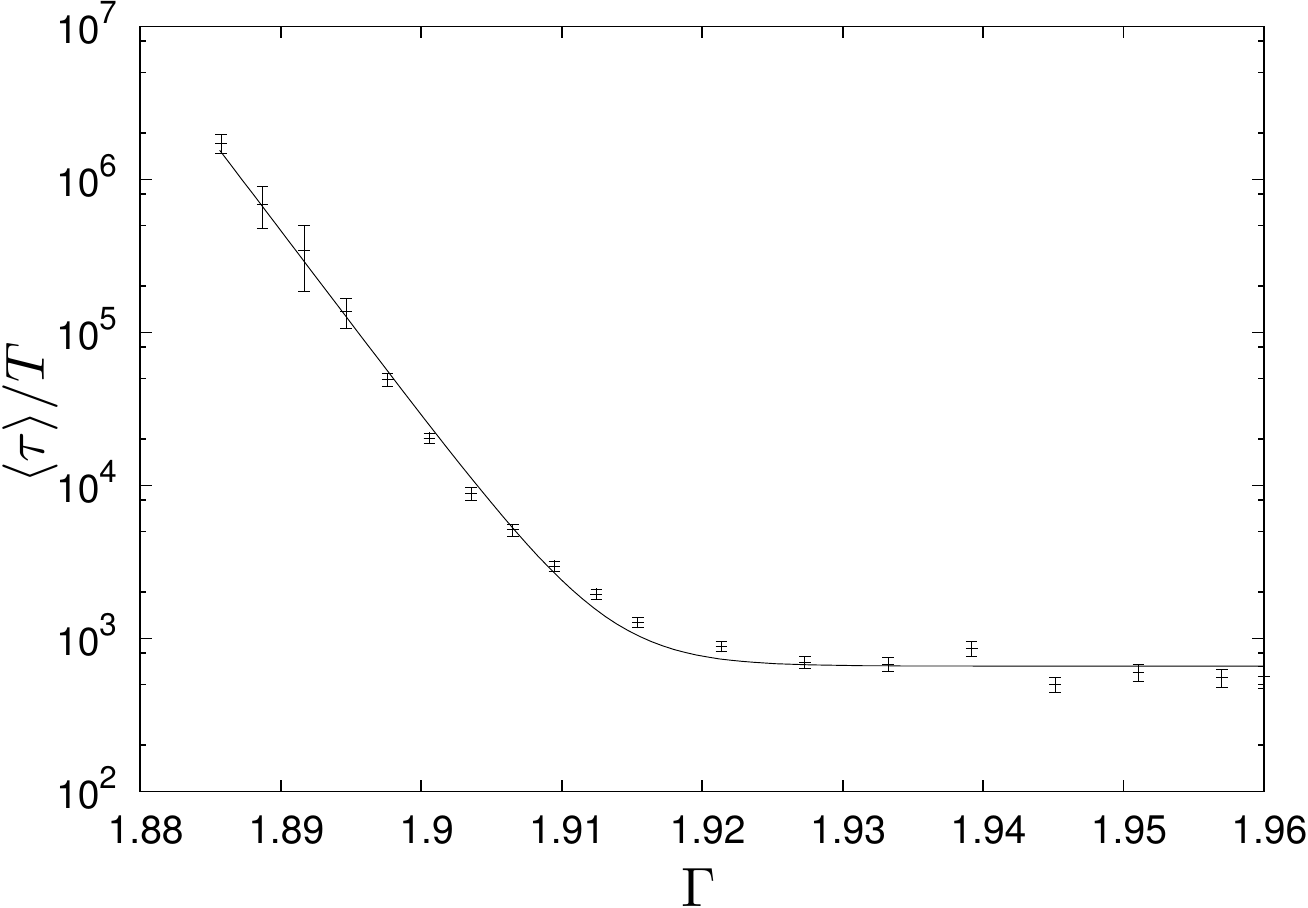}
\caption{Average transition time $\langle \tau \rangle$ to the absorbing state as a function of the dimensionless acceleration $\Gamma$, obtained in simulations.
The average and errorbars are computed from $N_r$ independent realizations.  The solid line corresponds to an exponential fit $
\langle \tau \rangle/T = a+\exp[{ c (\Gamma_0-\Gamma)]}$ with 
$a = 660\pm 43$, $c= 279\pm 7$, $\Gamma_0=1.94\pm0.01$.
The number of particles is fixed to $N=5160$.}
\label{fig.tauvsA} 
\end{figure}

Experimentally, it is difficult to vary the number of particles as it is done in the simulations. This would require to open and close the experimental setup each time $N$ is changed, affecting the reproducibility. Instead, we decided to vary the vibration amplitude $A$, and thus the acceleration $\Gamma$, while keeping fixed the frequency and the number of particles. Simulations with homogeneous initial conditions, were the number of particles is fixed but the amplitude is varied, were also performed, showing the same phenomenology as when the number of particles is varied (Fig. \ref{fig.tauvsA}). Here, a rapid increase of the transition times is also obtained although the results are better fitted to a single exponential function as compared to a double exponential when the number of particles was varied. In Sec.\ \ref{sec.espNA} we will analyze the relation between these two control parameters.

\subsection{Quasi-absorbing state: Alternating transitions between fluidized and synchronized states}

In the current setup, the system is first vibrated a low acceleration $\Gamma_{\rm f}= 2$, reaching a steady state in a fluidized regime. After, the vibration amplitude is rapidly changed to a final value $\Gamma$. For $1<\Gamma \lesssim 2.1$ the system reaches quickly a fluidized steady state and remains in this state as long as the experiment is run. When $\Gamma\gtrsim 2.1$, we observe that the system transits to the quasi-absorbing state with all grains moving vertically showing a small residual horizontal motion. This transition can occur quickly or after a transition time depending on the value of $\Gamma$. In the quasi-absorbing state particles move collectively synchronized with the cell's vertical motion.  Similarly as done in \cite{explosion}, this collective motion can be inferred from the acceleration signal or simply be visualized by acquiring images from a side view at a frame rate slightly different to the driving frequency. 

In contrast to simulations, this is a quasi-absorbing state because the system eventually can escape from it due to experimental defects. 
The described collective motion is not purely vertical: the bottom and top glass plates are not perfectly flat and a slow drift is observed in which grains move to the center. Then, suddenly, when the density in the central region reaches a critical value an energy burst event takes place: a pair of grains have an oblique collision that transforms the energy stored in the vertical motion to the horizontal degrees of freedom, triggering a chain reaction that can ultimately fluidize completely the system, similarly to what occurs in a bi-disperse mixture of equal size light and heavy particles \cite{explosion,explosion2}. By improving the quality of the experimental setup, the drift velocity can be reduced and therefore the residence time in the absorbing state can be increased up to $\approx 20$ s $= 1400T$. Also, there is a small random horizontal motion originated in  imperfections as particle non-sphericity, particle and wall surface roughness and external noise due to vibrations.

\begin{figure}[t!]
\includegraphics[width=.9\columnwidth]{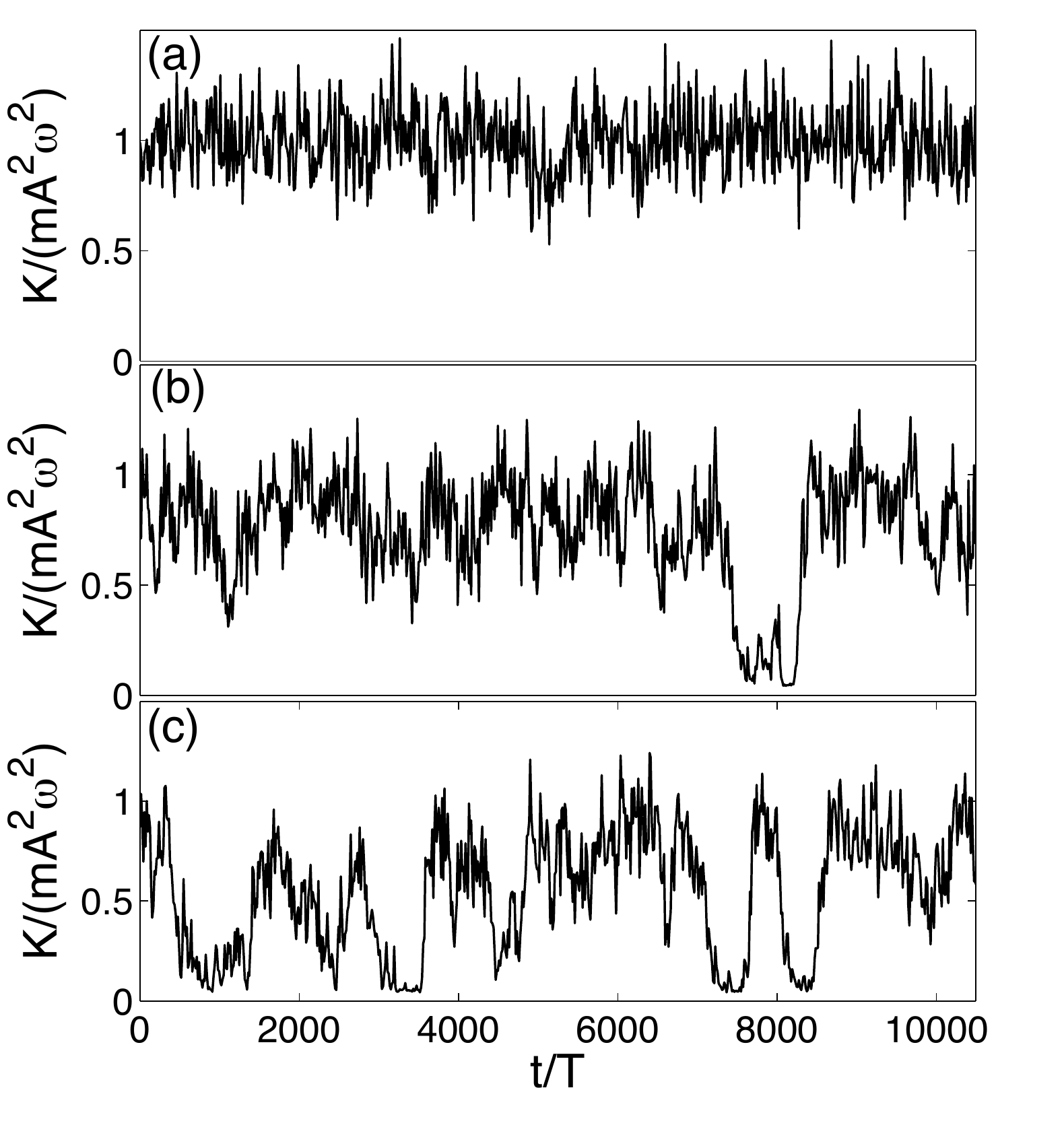}
\caption{Normalized horizontal kinetic energy per particle $K$ versus time for three driving accelerations, $\Gamma = 2.05 \pm 0.01$ (a), $\Gamma = 2.22\pm 0.03$ (b) and $\Gamma = 2.24 \pm 0.04$ (c), corresponding to a pure fluidized state (a) and alternations between fluidized and absorbing states (b,c). The normalization is done by the characteristic energy scale of one particle of mass $m$ with speed $A\omega$.}
\label{fig.expseries}
\end{figure}

The presence of the energy burst events, however, has a practical beneficial effect as it resets the system after reaching the absorbing state. This allows us to collect more statistics on the time the system remains in the fluidized regime without having to stop and restart the experiment. Typical horizontal kinetic energy time series are shown in Fig.~\ref{fig.expseries}. Here, kinetic energy is computed from particles velocities obtained from images, as the one shown in Fig. \ref{Exp_Cell}c. In Fig.~\ref{fig.expseries}(a) we show an example for the fluidized state. In Figs. \ref{fig.expseries}(b) and \ref{fig.expseries}(c) we present the alternating transitions to the absorbing state, followed by energy burst events.
As in the simulations, the fluid state is metastable and there is a large variability of the transition times to the absorbing state. As evident from Figs. \ref{fig.expseries}(b) and \ref{fig.expseries}(c), the transition times become shorter as $\Gamma$ is increased.

\begin{figure}[t!]
\includegraphics[width=.85\columnwidth]{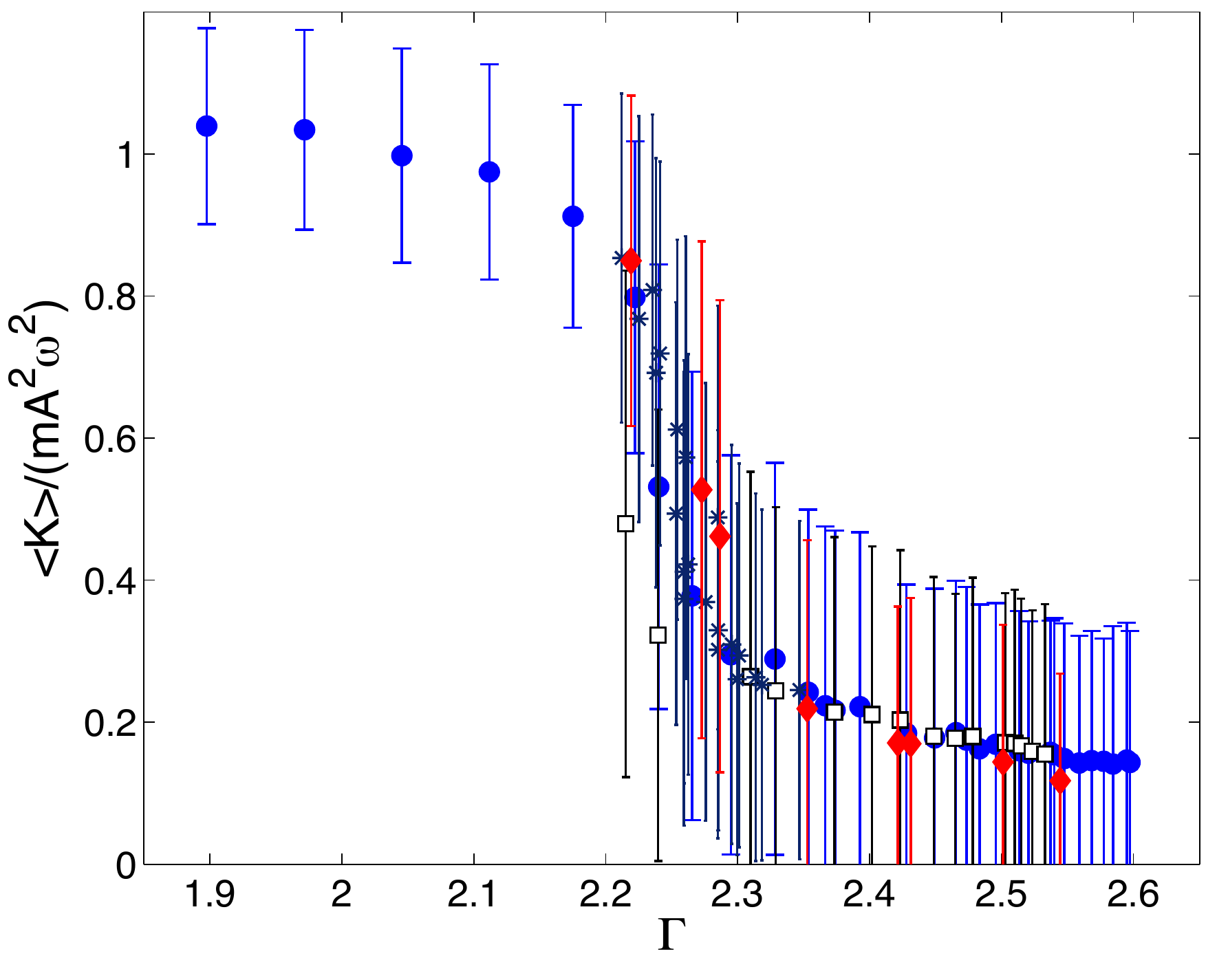}
\caption{(Color online) Time averaged normalized horizontal kinetic energy per particle $K$ versus $\Gamma$.  Data from four different $\Gamma$ ramps are presented, each one with a different symbol. Errorbars are computed from the standard deviation of each $K$ time series. In this case, the time series were of $300$\ s duration.}
\label{fig.K_vs_Gamma}
\end{figure}

The transition from the fluidized to quasi-absorbing state is clearly observed when considering the time average horizontal kinetic energy as a function of $\Gamma$, as presented in Fig. \ref{fig.K_vs_Gamma}. In this figure, four independent realizations are presented. For each one, a ramp in $\Gamma$ has been performed as described before. For the lower $\Gamma$ values, $\langle K \rangle / [m(A\omega)^2] \approx 1$, meaning that the average kinetic  energy scales well with $m(A\omega)^2$, were $m$ is the particle mass and $A\omega$ the maximum vibration velocity that fixes the characteristic energy scale. For $\Gamma \approx 2.2$, $\langle K\rangle$ decreases abruptly first and then more slowly for higher vibration amplitudes. As mentioned before, close to the transition the granular system is in a metastable state and the  transition times have a large variability, which explains the different $\langle K \rangle$ values obtained for different realizations in the vicinity of $\Gamma = 2.2$. This abrupt signature however does not contradict what we stated before, namely that the transition to the quasi-absorbing state is observed for $\Gamma \gtrsim 2.1$, but near this particular value the transition times are much longer and with a larger variability, as in simulations and as we will show below from experimental data acquired for longer times.

\begin{figure}[t!]
\includegraphics[width=.9\columnwidth]{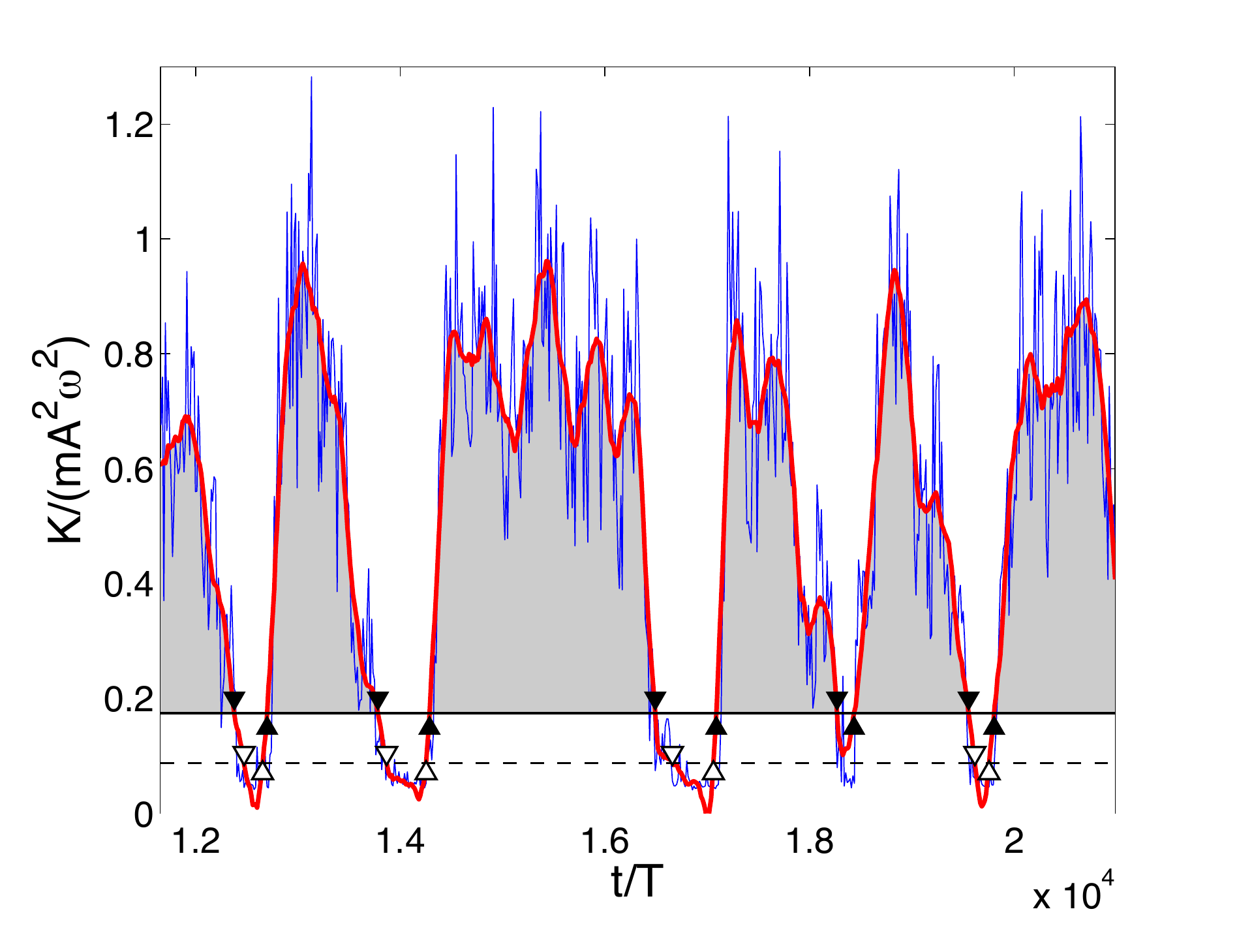}
\caption{(Color online) Example of fluidization time measurement from horizontal kinetic energy time series, $\Gamma = 2.24 \pm 0.04$. The original signal (blue, thin line) is smoothed (red, thick line). The horizontal dashed and continuous lines show two possible threshold values, $K_c = K_{\rm int}$ and $K_c = 2K_{\rm int}$ respectively, where the transition kinetic energy $K_{\rm int}$ is defined in the Appendix \ref{app.threshold}. Up and down triangles show times where a fluidization stage starts and ends, which define $\tau$ as the time lapse between theses events, represented by the gray areas below the filtered signal.}
\label{fig.ejemploKc}
\end{figure}

In Fig. \ref{fig.ejemploKc} we present an example of signal analysis in order to show how the fluidization times are computed. The original signal (thin continuous line) is smoothed using a Savitzky-Golay filter (thick continuous line). The filter's window size is $6.2$ s $= 434T$ (31 points at 5 Hz acquisition rate). This smoothing allows us to discard artificial short fluidization times induced by noise in the original signal. The fluidization times are then computed by fixing a threshold $K_c$. The determination of the threshold is described in Appendix \ref{app.threshold}. In this figure two possibilities are shown, $K_c = K_{\rm int}$ and $K_c = 2K_{\rm int}$, where $K_{\rm int}$ is the kinetic energy that divides the lower values characteristic of the quasi-absorbing state and the large values of the fluidized states. There are some synchronization events (absorbing state) that can be missed for the first, lowest threshold, as shown in this figure. The second criterium is more robust, being the final result, the analytical dependence of $\langle \tau \rangle$ on $\Gamma$, not modified by the particular choice we make. 

\subsection{Fluidization time analysis}

Using the kinetic energy times series it is possible to identify the events when the system transits to the absorbing state or when it refluidizes. However, the obtention is the kinetic energy is limited to relatively short time series due to limitations on the image acquisition process. However, as described in the Appendix \ref{app.pid} it is possible to create synthetic kinetic energy series using the information of the shaker's voltage input in a standard PID feedback controller that keeps the cell's acceleration constant. This PID control allows us to measure long voltage time series, much less expensively with respect to hard drive space and data analysis. In fact, the image processing is limited by the camera's hard drive memory and the time resolution one needs for measuring velocity (with two images obtained at a time $\delta t $ apart) and the time resolution needed for describing appropriately the system dynamics. In order to take the maximum advantage of the camera's capabilities, pairs of images acquired $\delta t~=~Ê4$ ms apart were in turn acquired at a lower frequency rate, separated in time by $\Delta t = 200$ ms. This constrains the duration of the image acquisition to $300$~s per video, corresponding to $21000$ cell  oscillations. For the voltage signal acquisitions the hard drive limitations are much less constrained and we acquire 10 times longer times series, of maximum duration $3000$ s.

\begin{figure}[t!]
\includegraphics[width=.9\columnwidth]{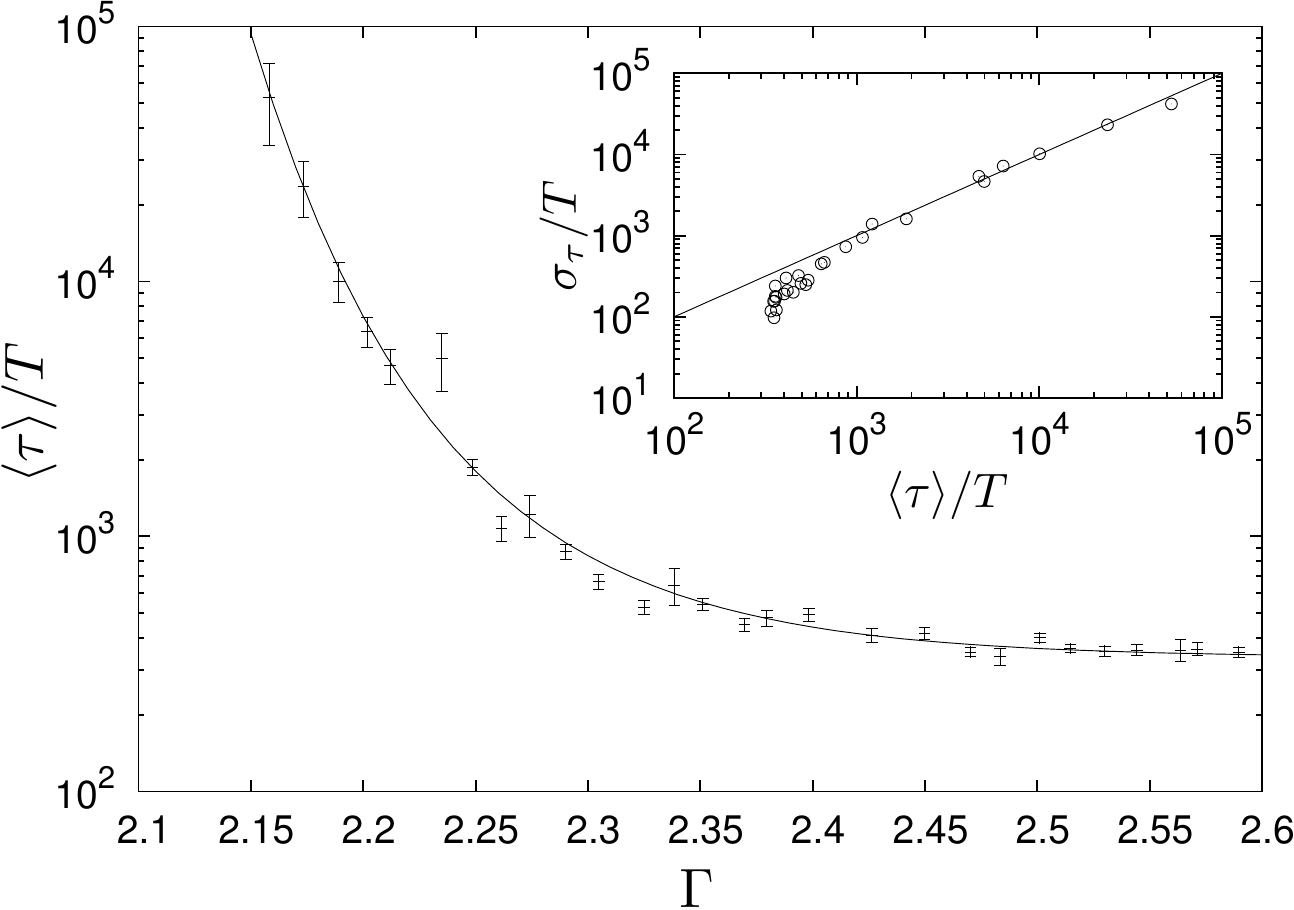}
\caption{Average metastable fluidization time $\langle \tau \rangle$ versus $\Gamma$ measured in experiments. The solid line is a double exponential fit. The inset shows the standard deviation $\sigma_\tau$ versus average fluidization time $\langle \tau \rangle$. The continuous line shows the Poissonian process result $\sigma_\tau = \langle \tau \rangle$. }
\label{fig.trtimes.expbc}
\end{figure}

For each value of the vibration acceleration the set of times $\tau$ the system remains in the fluid phase, which are computed using the threshold defined in the previous section. Those times are obtained from the horizontal kinetic energy and voltage time series, giving consistent results. However, being the voltage signals of longer duration we find larger $\tau$ values than from the kinetic energy data. Details are given in the Appendix \ref{app.pid}. 
In Fig. \ref{fig.trtimes.expbc} we  show average values $\langle \tau \rangle$ as function of $\Gamma$. Values that are $\gtrsim 5000T$ are solely obtained with the voltage time analysis, whereas values $\lesssim 5000T$ are obtained from both kinetic energy and voltage time measurements. Averages are obtained by dividing the data using windows $\Gamma \in [\Gamma_i,\Gamma_{i+1}]$, with $\Delta \Gamma = \Gamma_{i+1} - \Gamma_i =  [\max(\Gamma) - \min(\Gamma)]/N_w$ and $N_w=30$. Again, the conclusions that we obtain do not depend on the value of $N_w$. As for simulations, for the errorbars we use the standard error definition $\sigma_\tau/\sqrt{N_r}$, where $\sigma_\tau$ is the standard deviation of the $N_r$ measurements of the fluidization time in a given window of $\Gamma$. 

We observe that, as in simulations,  the average fluidization time $\langle\tau\rangle$ presents a rapid increase for a small variation of the control parameter:  a $10\%$ reduction in $\Gamma$ produces a change in a factor of $150$ in $\langle \tau \rangle$. Here the increase is obtained for a reduction of the control parameter, contrary to what occurs in the simulations, but this difference is irrelevant for the comparison.
In Fig. \ref{fig.trtimes.expbc} we also present a fit with a double exponential law
\begin{equation}
\langle \tau \rangle/T  =  \exp [a + \exp[c (\Gamma_o - \Gamma)]],
\end{equation}
with $ a = 5.8 \pm 0.1$, $ c = 12.1 \pm 1.4$ and $ \Gamma_o = 2.29 \pm 0.02$. For both simulations and experiments, this law fits better our results than a single exponential law and a critical divergence. The initial nucleus study that we presented before rules out the possibility of the critical divergence. Interestingly, this mean lifetime double exponential law has also been found in the transition to turbulence for both Taylor-Couette and pipe flows \cite{Avila2011,Dauchot2012}. In simulations we found both a single exponential and a double exponential, depending on which control parameter was varied. More statistics and longer experiments are needed to be able to give a definitive answer on how the average transition times increase.

Finally, in the inset of Fig. \ref{fig.trtimes.expbc} we present the standard deviation $\sigma_\tau$ as function of the average fluidization time $\langle \tau \rangle$ obtained  experimentally. The solid line shows the Poissonian result $\sigma_\tau = \langle \tau \rangle$. As for simulations, the agreement is quite good, except at the lower average times where $\langle \tau \rangle$ tends to saturate.

\section{Continuous model}  \label{sec.landau}

To describe the observed transition we use a Ginzburg-Landau approach. The order parameter is the horizontal kinetic energy $K$ and no other  field is identified to be relevant for the dynamics close to the transition. Density is not observed to couple noticeably with $K$. Momentum density and vertical kinetic energy are non-conserved fields having fast dynamics and, therefore,  are not important in the description of the slow transition process.

Both at the metastable state (with $K$ finite) and at the absorbing state there is no observable residual dynamics. It is then plausible to model the dynamics of the order parameter as purely variational. As $K$ is non-conserved, model $A$ is appropriate for the deterministic dynamics~\cite{HalperinHohenberg,Sancho}
\beq
\derpar{K}{t} = -\lambda \frac{\delta F}{\delta K} , \label{eq.variational}
\eeq
where $\lambda>0$ is a transport coefficient and $F$ is the variational functional, also called non-equilibrium free energy. 

\begin{figure}[t!]
\includegraphics[width=.85\columnwidth]{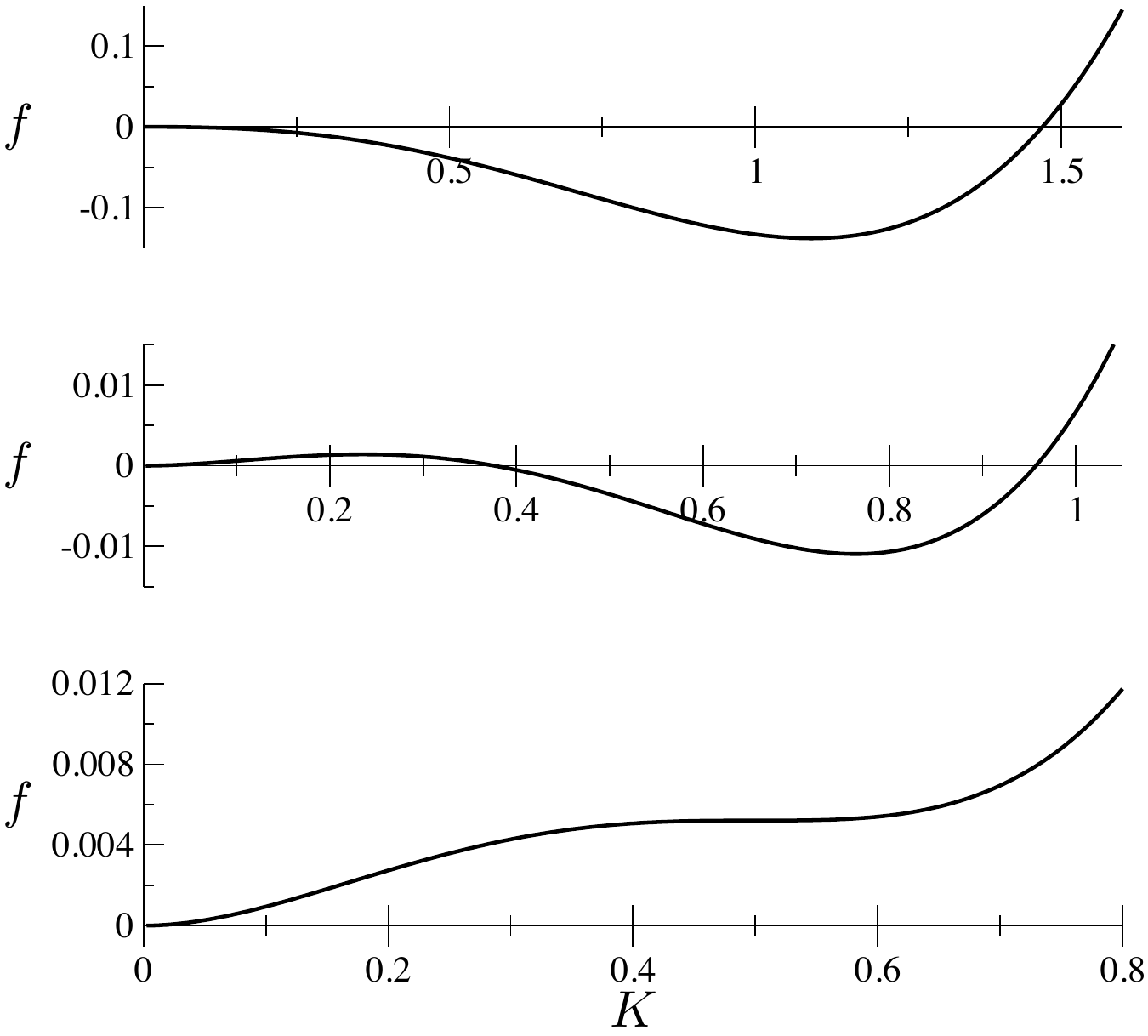}
\caption{Landau free energy as a function of the horizontal kinetic energy. 
Top: Case with $\epsilon<0$ ($\epsilon=-0.1$). Only the state with $K\sim 1$ is an stable equilibrium.
Middle: Case with $\epsilon>0$ ($\epsilon=0.18$). Two stable equilibrium states are found with $K=0$ and $K= K_1$. The barrier is at $K\sim \epsilon$. Bottom: Case with $\epsilon=1/4$. The equilibrium at $K=K_1$ becomes unstable via a saddle point.}
\label{fig.freeenergy}
\end{figure}

Using the Ginzburg approach, the free energy functional is written as
\beq
F = \int d^2r \left [ f(K(\Vec r)) + \mu (\nabla K)^2\right],
\eeq
that is, the sum of  a local free energy plus a term that penalizes high inhomogeneities and will be responsible of an effective surface tension between the two phases. 
To build up $f$ we consider first the deterministic  dynamics.
As the free energy enters as a derivative, any additive constant is irrelevant and we fix $f(0)=0$.
The order parameter must remain positive, therefore $df/dK|_{K=0}\leq0$.
The  variational dynamics implies that the system is attracted to the local minima of $f$.
The fluid state is always a local minimum but varying the control parameter (the number of particles) it is possible to make that $K=0$ becomes also a minimum. 
We call $\epsilon$ the difference of the control parameter to the critical one, such that for $\epsilon>0$, $K=0$ is stable and therefore  $df/dK|_{K=0}=0$.
Using the assumptions of Landau theory that the free energy should be analytic on the control and order parameters, $df/dK|_{K=0}=0$ for all values of $\epsilon$. Only the concavity at $K=0$ changes with $\epsilon$. Finally, for $|\epsilon| \ll 1$ a second minimum at $K>0$ should be present. With these restrictions, the simplest expression for the local free energy is $f=a\epsilon K^2 -b K^3+c K^4$ with $a$, $b$, and $c$ positive constants.

Defining appropriate units of kinetic energy and time and rescaling the bifurcation parameter $\epsilon$, Eq. \reff{eq.variational} reduces to 
\beq
\derpar{K}{t} = -\epsilon K +K^2-K^3 + D \nabla^2 K. \label{eq.wonoise}
\eeq
The coefficient $D$ could also be absorbed by a proper change of length units, but we prefer to leave it as a free parameter and fix length units using the box size instead. The corresponding free energy is
\beq
f= \epsilon K^2/2-K^3/3+K^4/4 \label{eq.freeenergy}. 
\eeq
As mentioned before, $\epsilon$ is proportional to the distance of the number of particles to its critical value. The relation of $\epsilon$ with amplitude in the experimental case will be elucidated in Sec. \ref{sec.numericalEDP}.

\begin{figure}[t!]
\includegraphics[width=.85\columnwidth]{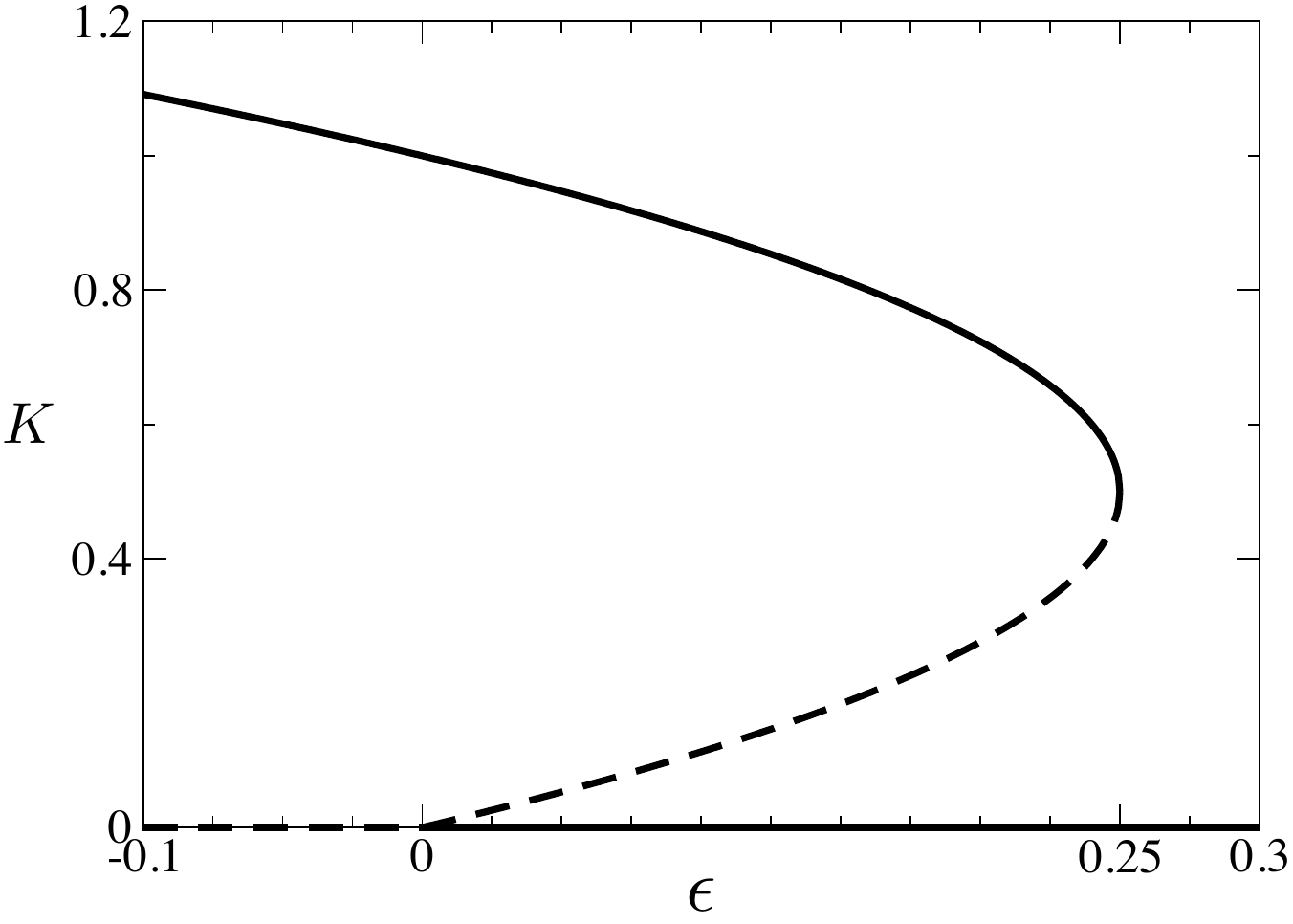}
\caption{Bifurcation diagram of the continuous model~\reff{eq.wonoise}. Solid and dashed lines represent stable and unstable equilibria, respectively.
Note that only positive values for the order parameter are possible. The critical points are located at $\epsilon=0$ and $\epsilon=1/4$.}
\label{fig.bifurcationEDP}
\end{figure}

The homogeneous stationary states can be obtained directly from~\reff{eq.wonoise} for which we give the exact expressions and their approximate values in the limit $\epsilon\ll1$. First, $K_0=0$  and  $K_1=(1+\sqrt{1-4\epsilon})/2\approx 1$ are always equilibrium states; the first is stable only if $\epsilon>0$ and the second is stable for $|\epsilon| \ll 1$. When $\epsilon>0$ there is another stationary state $K_2=(1-\sqrt{1-4\epsilon})/2\approx \epsilon$, which is unstable. 
The free energy at the stationary states are: $f(K=K_0)=0$, $f(K=K_1)\approx -1/12$, and $f(K=K_2)\approx\epsilon^3/6$. Schematic representations of the free energy are shown in Fig. \ref{fig.freeenergy} and the bifurcation diagram of the model is presented in Fig.~\ref{fig.bifurcationEDP}.
The analytic form of $f$ introduces an extra transition at $\epsilon=1/4$, where $K_1$ becomes unstable via a collision with $K_2$. This second transition point is a result of our model and will play an important role in the analysis presented in Sec. \ref{sec.numericalEDP} as was noted previously in the context of population dynamics \cite{Meerson}.

As shown in the figures, when $\epsilon>0$, the state with $K=K_1$ is stable. If normal, additive, fluctuations were added, as the free energy in this state is the smallest, it would become the global equilibrium. However, it is observed that $K=K_0$ is an absorbing state, without fluctuations.
We model this fluctuating dynamics introducing a noise term in \reff{eq.wonoise} which acts as a source of horizontal kinetic energy. This energy can only come from the vertical motion of the grains and the transfer efficiency is proportional to the collision rate which, in turn, is proportional to $\sqrt{K}$. The fluctuating equation then reads
\beq
\derpar{K}{t} = -\epsilon K +K^2-K^3 + D\nabla^2 K+\sqrt{\Gamma_{\rm n} K} \, \eta(\Vec r, t), \label{eq.wnoise}
\eeq
where $\Gamma_{\rm n}$ measures the intensity of the noise and $\eta$ is assumed to be a white noise with correlation 
\beq
\langle \eta(\Vec r,t) \eta(\Vec r',t') =\delta(\Vec r -\Vec r') \delta(t-t').
\eeq
The noise is multiplicative and the equation should be interpreted according to Ito as it corresponds to an internal noise \cite{vanKampen,Gardiner}. 
The noise in the simulations and experiments is due to the discreteness of the granular system with complex (sometimes chaotic) dyanmics.

The model \reff{eq.wnoise} is an analog to models for population dynamics with absorbing (extinction) states that have been studied in one dimension \cite{Elgart,Meerson}. Those models start with a discrete populations that move on lattice sites in continuous time and the analysis is performed in the case of large but finite average populations. Our case is continuous in space and the order parameter is continuous by essence although the number of particles is discrete. Despite these differences, it is expected that the analysis for the one dimensional case is relevant here. 
It was shown that regardless of the sign of $\epsilon$ the absorbing state is always reached but, when $\epsilon<0$, the transition time grows exponentially with the system size as the most probable transition path is that the energy homogeneously decreases to zero  \cite{Elgart,Meerson}. 
The case when $\epsilon>0$, in what is called  the Allee effect, was studied in detail in Ref. \cite{Meerson}. It is found that  the transition to the absorbing state takes place via nucleation with a critical nucleus. Close to $\epsilon=1/4$ the critical nuclei are system size-independent and the transition time is dominated by the creation the critical nucleus, for which the energy the order parameter must overcome a small barrier located at a finite $K$ (see Fig. \ref{fig.freeenergy}, bottom). Therefore, for the purpose of computing the transition time the noise term can be approximated by an additive one with $K\approx K_1$ and usual homogeneous nucleation theory can be used \cite{Langer,HNT1,HNT2}. Far from $\epsilon=1/4$ the transition is also via nucleation, but the  transition times grow exponentially with the system size. Finally, much less is known in the  weak Allee effect, for $0<\epsilon\ll 1/4$.

\begin{figure}[t!]
\includegraphics[width=.85\columnwidth]{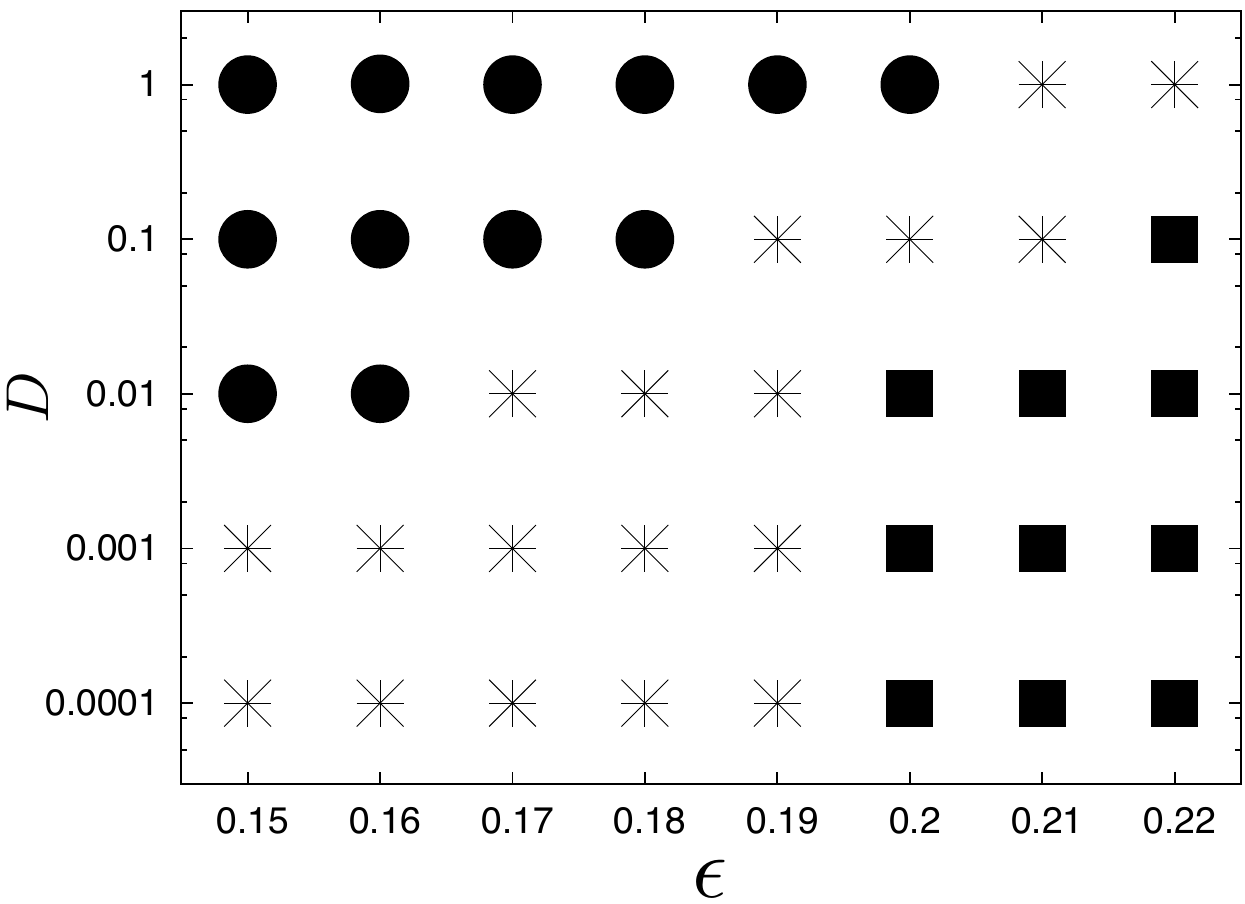}
\caption{Characterization of the final state in  $\epsilon-D$ parameter space of numerical solution of Eq. \reff{eq.wnoise}. At high $D$ and small $\epsilon$ values the system remains in the fluid state (solid circles), at low $D$ and large $\epsilon$ values there is a transition to the absorbing state (solid squares) and in the intermediate region, the final state depends on the realization of the noise (stars). For $\epsilon\leq 0.14$ no transition to the absorbing state is observed.}
\label{fig.spontaneoustransitionEDP}
\end{figure}

\subsection{Numerical analysis} \label{sec.numericalEDP}

Eq. \reff{eq.wnoise} is solved numerically using a pseudo spectral method. A square box of unit size and periodic boundary conditions is discretized using $100\times 100$ grid points. Time is discretized with a step $\Delta t =0.01$. The Laplacian term is integrated in time using the Crank-Nicolson method in Fourier space and the local and noise terms are integrated in real space using Euler's method. The initial condition is a fluidized state, that is, all points are set to $K_1$ for the specific value of $\epsilon$. Under these conditions,  the system dynamics is described by only three dimensionless control parameters: $\epsilon$, $D$, and $\Gamma_{\rm n}$.

Eq. \reff{eq.wnoise} is solved for different control parameters. As expected, when $\epsilon<0$, the fluidized initial condition is stable.
When $\epsilon$ takes small positive values two regimes were observed depending on the noise intensity. For large noise intensities ($\Gamma_{\rm n}\gtrsim10^{-3}$, with a precise threshold that depends on $\epsilon$ and $D$), the system transits to the absorbing state simultaneously in all the space, albeit with large fluctuations. The evolution is as if there were many simultaneous nucleation sites with no appreciable metastability. On the contrary, if noise intensity is small, the system remained in the vicinity of $K_1$ as long as the simulations took place ($T_{\rm wait}=10^5$) and no transition was observed. 
That is, for small positive $\epsilon$ no transition via a single nucleus is observed.
This result indicates that the effective noise intensity in simulations and experiments is small.

\begin{figure}[t!]
\includegraphics[width=.9\columnwidth]{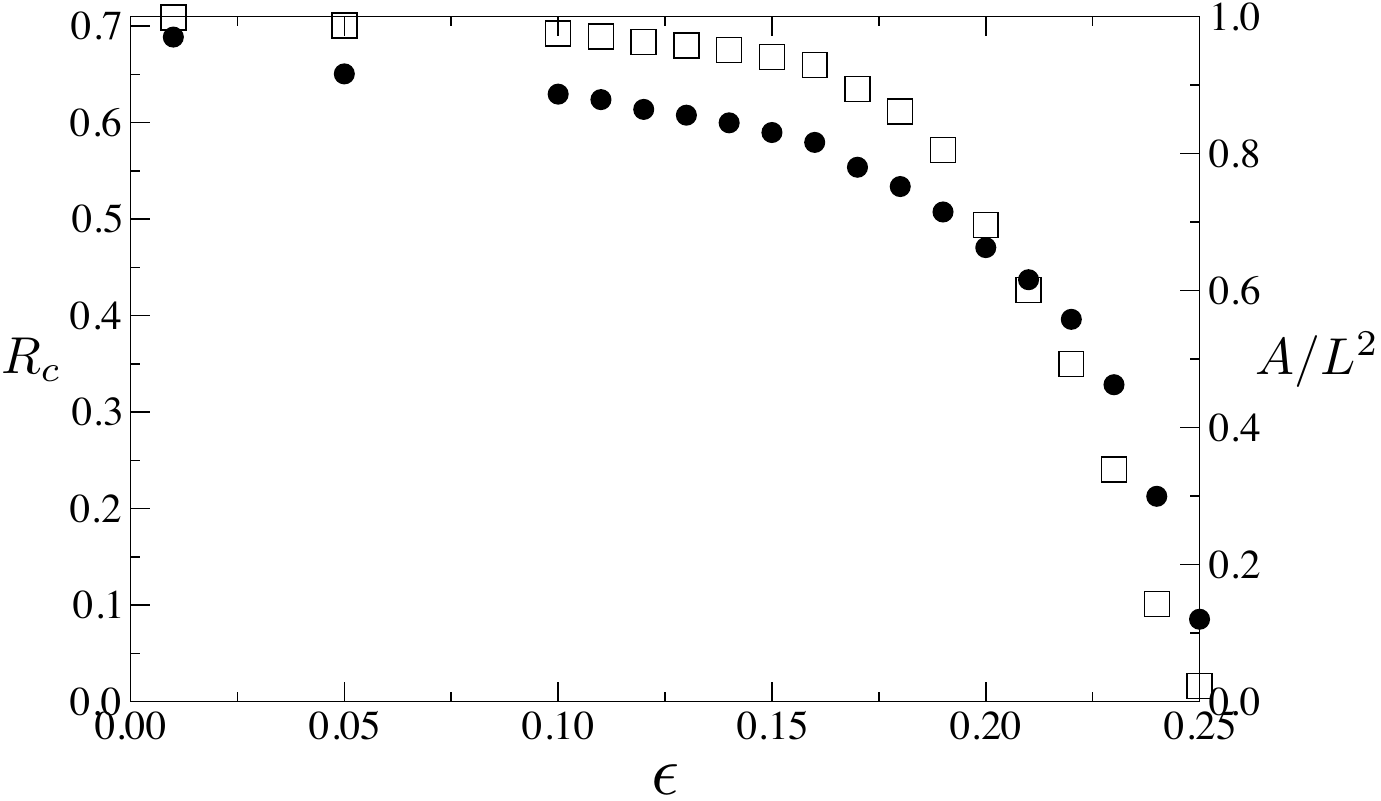}
\caption{Radius (solid circles, left axis) and relative area (open squares, right axis) of the critical nuclei as a function of the bifurcation parameter $\epsilon$ obtained in the continuous model \reff{eq.wnoise} on a square box of unit size. The maximum nucleus radius is $R_{\rm max}=\sqrt{2}/2\approx 0.71$; beyond that value the nucleus covers the complete system. The diffusion coefficient is $D=0.05$ and the noise intensity is $\Gamma_{\rm n}=0$.}
\label{fig.nucleusEDP}
\end{figure}

To increase further the possibility to observe the transition via nucleation, $\epsilon$ is allowed to take larger positive values. 
Fixing the noise intensity  $\Gamma_{\rm n}=2.4\times 10^{-4}$ a series of simulations varying $\epsilon$ and $D$ were performed for a long waiting time, $T_{\rm wait}=10^5$. Transitions are observed only when $\epsilon$ takes large values, close to $1/4$, when $K_1$ becomes unstable. The parameter space diagram is shown in Fig.~\ref{fig.spontaneoustransitionEDP}.

As critical nuclei take extremely long times to spontaneously form when $\epsilon$ is small, simulations with an initial nucleus are performed. A circular nucleus  of radius $R$ in the absorbing state ($K=K_0$) is initially created, while the rest of the system is in the fluid state ($K=K_1(\epsilon)$).  Equation~\reff{eq.wnoise} is solved until either the complete system reaches the absorbing state or it is fluidized entirely. In this case, the noise intensity is set to zero as fluctuations do not change appreciably the outcomes. This procedure allows us to identify the critical nucleus radius $R_c$ separating both possible outcomes. Fig.~\ref{fig.nucleusEDP}, which should be compared with the simulation results in Fig. \ref{fig.criticalradius}, presents the critical nucleus radii as well as the area occupied by the critical nucleus relative to the box area. It is obtained that the critical nuclei are of macroscopic size (that is, a finite fraction of the whole system) except very close to $\epsilon=1/4$, in agreement with the theoretical analysis in one dimension \cite{Meerson}. Even at $\epsilon=0.22$ the nucleus has to occupy more than half of the system to effectively drag the complete system to the absorbing state.
A second series of simulations with a coarser grid ($50\times 50$) was performed with equivalent results, showing that the obtained critical nuclei are not dependent on the discretization.

The comparison of the model with the simulations and experiments is qualitatively correct, showing that the dynamics is described by a free energy which is qualitatively similar to \reff{eq.freeenergy}. The macroscopic size of the critical nuclei, that is completely different to the situation in homogeneous nucleation in equilibrium \cite{Langer,HNT1,HNT2}, is due to the multiplicative noise \cite{Elgart,Meerson}. Indeed, at $\epsilon=0$ the absorbing state becomes the global equilibrium but the free energy difference between the absorbing state and the fluid state remains finite ($\Delta f =1/12$). Therefore the formation of a nucleus has a volumetric energy cost and not only a surface energy cost as it would be in the case of equilibrium at the coexistence curve. The free energy difference remains finite up to $\epsilon=0.22$ and it is only negative (favorable to spontaneous transitions in an equilibrium case) in the narrow range $0.22\leq\epsilon\leq0.25$, corresponding to the strong Allee effect where the critical nucleus become of finite size \cite{Meerson}. These hand waving argument should be taken only as a qualitative explanation but that cannot be be considered as the full explanation: because when the noise is multiplicative the stationary distributions are not given by a Boltzmann distribution and, therefore, the transition probabilities are not given simply by differences of free energy \cite{Gardiner,Sancho}. A detailed analysis in the context of population dynamics in 1D is performed in Refs. \cite{Elgart,Meerson}.

\subsection{Relation of $\epsilon$ with the control parameters} \label{sec.espNA}
The comparison with the simulations and the theoretical analysis suggest that not only the critical point at $\epsilon=0$ is relevant for the dynamics as it is unveiled looking at the critical nuclei, but also the critical point at $\epsilon=1/4$. Near it, the critical radii are small and, therefore, nuclei can be spontaneously created by fluctuations in finite time. This second critical point becomes evident in simulations of the continuous model when fluidized states are let to evolve spontaneously to the absorbing state. The simulations and the model can then be mapped in the following way: $\epsilon=0$ corresponds to $N=N_{\rm max}=11547$ and $\epsilon=1/4$ corresponds to $N\approx N_0\approx5130$.
Note, however, that the analytic expression \reff{eq.freeenergy} becomes only qualitatively correct  when applied for large values of $\epsilon$, as the proposed free energy is built as a power series.

The experimental control parameter is the vibration acceleration $\Gamma$, while the number of particles $N$ was kept constant. When mapping the experiment to the model it is expected that $\epsilon=\epsilon(\Gamma,N)$. Based on the same arguments as in the simulations, $\epsilon$ should vanish only when $N=N_{\rm max}$, that is $\epsilon(\Gamma,N_{\rm max})=0$ for a wide range of amplitudes. The observed transition of spontaneous transitions to the absorbing state implies that $\epsilon(\Gamma,N)$ crosses $1/4$ at $\Gamma\approx 2.3$. The complete mapping can only be obtained if the crossing points are found for different $N$.

\section{Conclusions and perspectives} \label{sec.conclusions}

We have considered a quasi two-dimensional granular system in a box that is vertically vibrated. This configuration presents an absorbing state with all grains bouncing in phase with the box and no horizontal motion. In molecular dynamics simulation this state is perfectly absorbing while in experiments there is a residual small horizontal motion that triggers rare chain reaction refluidization  events. These imperfections reset the system once in the absorbing state and allows us to make statistical analysis on the transition to the absorbing state. The transition is of first order, with long metastable times in an active fluidized state, which are Poisson distributed, consistent with a Markovian process for the relevant variable. In simulations it is shown that the transition is driven by nucleation with nuclei of irregular shapes.

Interestingly, the average mestastable times grow dramatically far below the critical density which corresponds to a packed monolayer. In practice, close to the critical density no simulations or experiments show spontaneous transitions to the absorbing state. With simulations transitions up to the critical density are possible by creating initial nuclei in the absorbing phase. However, the size of the critical nuclei occupy a large part of the system close to the critical density, which is extremely improbable to occur spontaneously by the internal noise.

The observed phenomenology is captured by a coarse-grained Ginzburg-Landau model with multiplicative noise, whose discrete one-dimensional stochastic version was previously analyzed in detail with similar results to our two-dimensional case. At the critical point of the maximal density, the transition can only occur with a critical nucleus that occupied the whole system, taking a time that grows exponentially with system size. At lower densities the critical nuclei are still of macroscopic size and only become of finite size, with a corresponding size-independent transition time, close to a second critical point where the active fluidized state becomes unstable via a saddle node bifurcation. This bifurcation is the one that is observed in the experiments and simulations for a critical density that depends on other parameters. The rapid increase of the metastable times are associated to the increasing stability of the active state.

\acknowledgments{
We thank M.G. Clerc and P. Cordero for valuable technical discussions. This research is supported by Fondecyt Grants No.~1120211 (N.M.), No.~1120775 (R.S.),  and No.~1100100 (R.S. and I.R.). B.N. and A.T. were respectively supported by ENS-Paris and the Chicago-Chile Material Collaboration NSF DMR-0807012.}

\appendix

\section{PID acceleration control. } \label{app.pid}

\begin{figure*}[t!]
\begin{center}
\includegraphics[width=1.95\columnwidth]{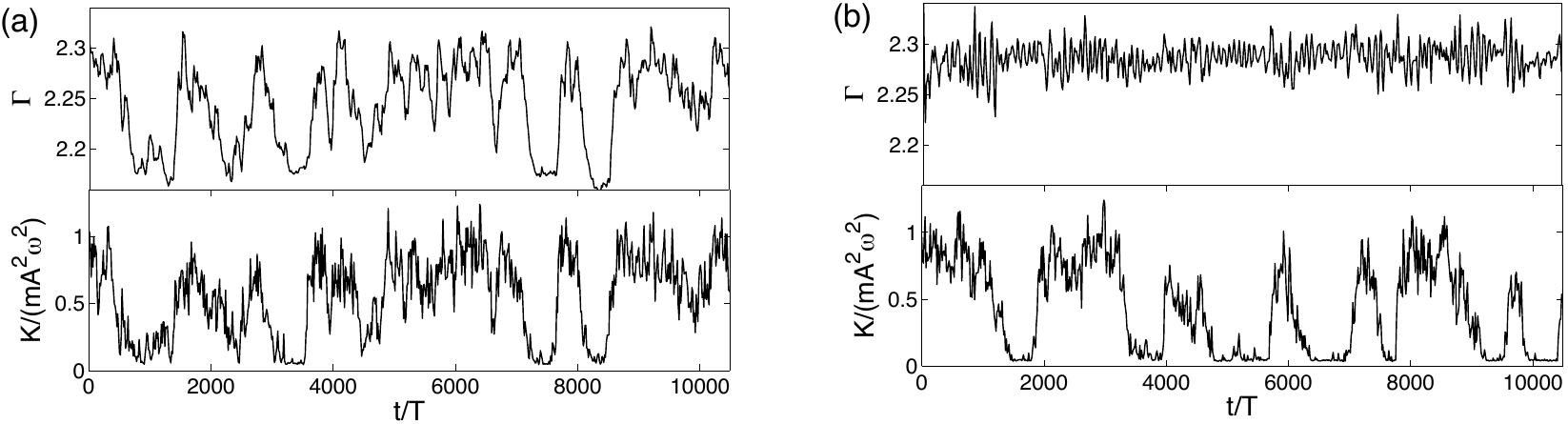}
\caption{Acceleration and normalized horizontal kinetic energy without (a) and with (b) PID control. The measured accelerations are $\Gamma = 2.24 \pm 0.04$ and $\Gamma = 2.29 \pm 0.02$ respectively. With no PID control, $\Gamma$ is clearly correlated to $K$: a transition to the absorbing state, indicated by a strong decrease in $K$, is concomitant with a small decrease in $\Gamma$, of about $5\%$.}
\label{K_PID}
\end{center}
\end{figure*}

\begin{figure}[t!]
\includegraphics[width=.95\columnwidth]{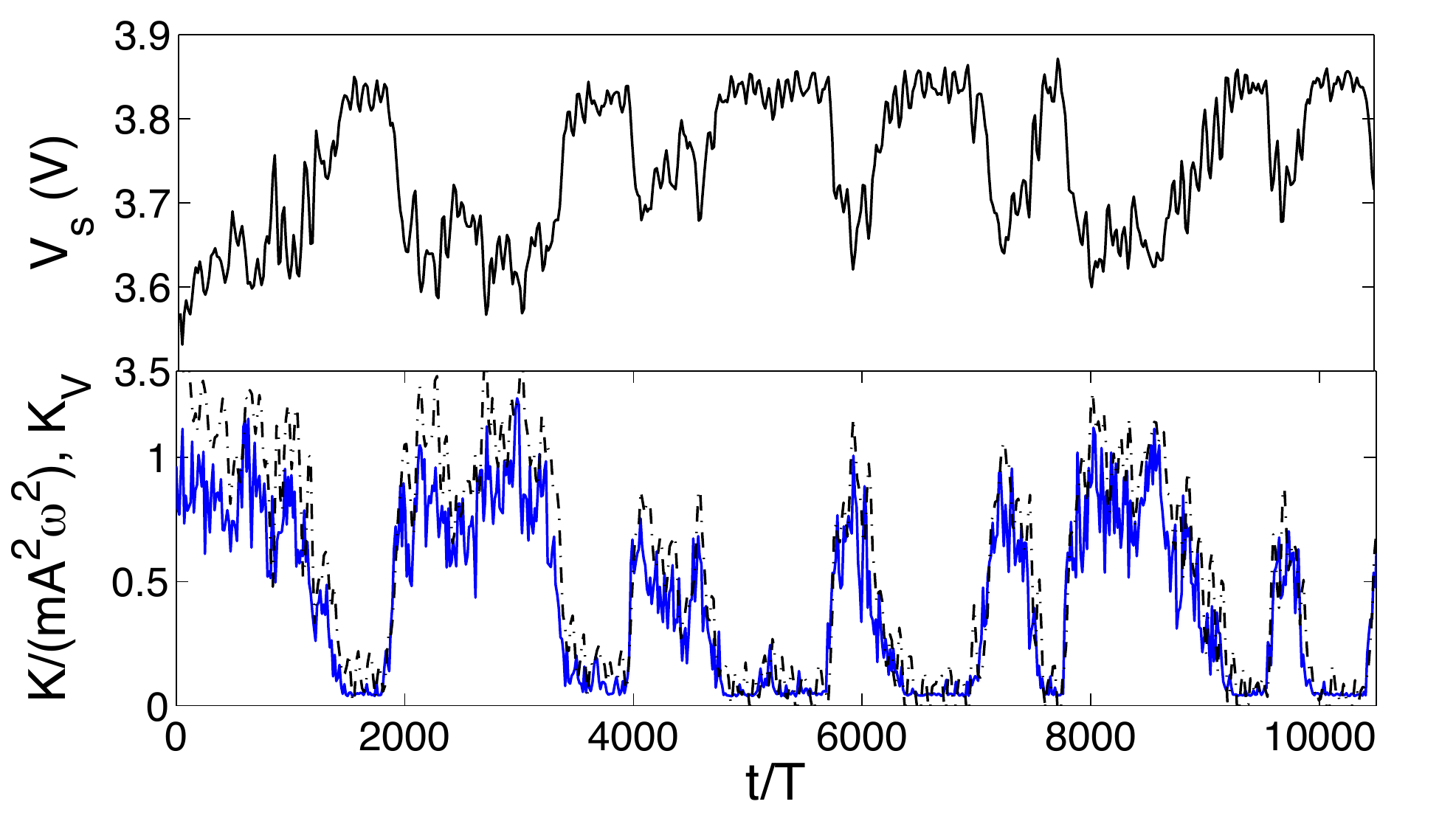}
\caption{(Color online) Driving voltage signal $V_{\rm s}$ (top) and horizontal kinetic energy (bottom) with PID control, $\Gamma = 2.29\pm 0.02$. $V_{\rm s}$ increases when the granular system transits from the fluidized to the quasi-absorbing state. The bottom panel shows the normalized horizontal kinetic energy $K/(mA^2\omega^2)$ obtained from particle tracking (continuous line) and the synthetic horizontal kinetic energy obtained from the voltage signal, $K_V = a_{\rm v} [\max(V_{\rm s})-V_{\rm s}]$ with $a_{\rm v}  = 5$ V$^{-1}$ (dashed line).}
\label{fig.voltagePID}
\end{figure}

An important experimental modification with respect to our previous studies \cite{explosion, Castillo} is the incorporation of a standard proportional-integral-derivative (PID) feedback controller for the driving acceleration. Without the PID control, the vertical vibration is actually performed at constant RMS power, provided by the power amplifier that feeds the electromechanical vibrator with an electrical AC signal. When the granular system transits from a fluidized to a quasi-absorbing state, its average dissipated power increases, implying that its average injected power also increases.  As the total power is kept constant, the excess of injected power to the granular system means less power for the cell's motion, implying a decrease of the amplitude of vibration and thus its acceleration $\Gamma$. An example of this is shown in Fig. \ref{K_PID}(a). In this case, when the system undergoes a transition from the fluidized to quasi-absorbing state, a decrease by a factor $10$ in $K$ is concomitant with a $5\%$ decrease in $\Gamma$. As shown in Fig. \ref{K_PID}(b), the PID control allows us to keep more constant the driving amplitude $A$, thus the driving acceleration $\Gamma$, by adjusting the voltage signal that is feed to the electromagnetic shaker. In this case, although some fluctuations in $\Gamma$ remain, there is a much lower correlation with $K$. As a quantitative measure of the effectiveness of the PID control, the standard deviation of $\Gamma$ is reduced by a factor of $2$ with the feedback control in place.

The PID control does not affect the observed phenomena. Indeed, the transitions between the fluidized and quasi-absorbing states are observed with and without the feedback control. Its advantage relies on the fact that the voltage signal used to feed the electromagnetic shaker is no longer constant. Because without the PID control the acceleration decreases when the transition to the quasi-absorbing state occurs, then, in order to keep the acceleration constant, a correct PID control should imply an increase of the voltage signal during this transition. This is indeed what occurs, as shown in Fig. \ref{fig.voltagePID}. This same figure demonstrates that from the voltage signal it is possible to construct a synthetic horizontal kinetic energy signal, $K_V = a_{\rm v} [\max(V_{\rm s})-V_{\rm s}]$, with $a_{\rm v}  = 5$ V$^{-1}$. The correlation between the real and constructed kinetic energies is indeed very good. Here, the constant $a_{\rm v}  = 5$ V$^{-1}$ is used in order to obtain such good agreement and is kept constant for all the subsequent voltage data analysis. 

\begin{figure*}[t!]
\includegraphics[width=1.95\columnwidth]{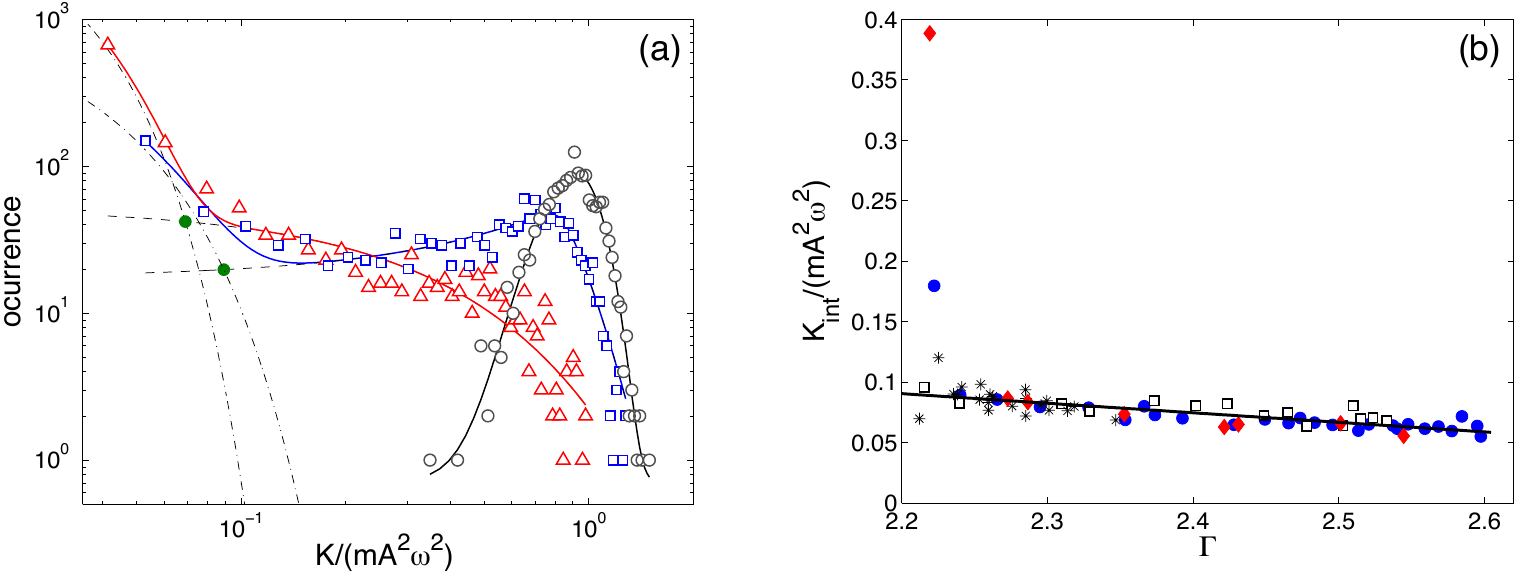}
\caption{(Color online) (a) Representative histograms of normalized horizontal kinetic energy, for $\Gamma = 2.18\pm0.02$ ($\circ$), $2.24 \pm 0.04$ ($\square$) and $2.45\pm0.03$ ($\triangle$). Solid lines correspond to either the fit of $F_1$ alone or the fit of the addition of $F_1$ and $F_2$. The dashed and dot-dashed lines correspond to the two fits $F_1$ and $F_2$. Solid circles ($\bullet$) correspond to the intersection values for the two largest $\Gamma$, $K_{\rm int}/(mA^2\omega^2) = 0.089$ and $0.069$, for which there are indeed transitions between the fluidized and quasi-absorbing states. (b) $K_{\rm int}/(mA^2\omega^2)$ versus $\Gamma$ for the same four realizations of Fig. \ref{fig.K_vs_Gamma}. The continuous line is a phenomenological linear fit $a\Gamma + b$, discarding the two largest $K_{\rm int}$ values.}
\label{fig.estadisticaK}
\end{figure*}

Finally, we stress that the same analysis can be performed on the acceleration signal (notice the good correlation between $\Gamma$ and $K$ in Fig. \ref{fig.voltagePID}(a)), but the changes in the driving voltage are larger than in acceleration, making transitions between the states more detectable. More importantly, the possibility of the existence of a critical point has to be considered. In this case, it is indeed much better to have a well controlled, constant driving parameter. As mentioned before, the PID control decreases the acceleration variations in about a factor~$2$. More importantly, these variations are no longer synchronized with the kinetic energy signal.

\section{Threshold determination for fluidization time measurements.} \label{app.threshold}

The main quantity to be measured is the fluidization time $\tau$, which is the time the granular system remains in the fluidized state before a transition to the quasi-absorbing state occurs. Our aim is to measure $\tau$ as function of $\Gamma$, both its average value and its fluctuations. This is done by defining a threshold for $K$ in order to differentiate both states. The fact that the transitions can occur alternatively allows us to analyze statistically this quantity.

Figure \ref{fig.estadisticaK}(a) shows representative histograms of normalized horizontal kinetic energy, for three different $\Gamma$. For the lowest acceleration the system remains in the fluidized state. In this case, the distribution is well fitted by a Gaussian, 
\begin{equation}
F_1(K_n) = A_1 e^{-\frac{(K_n - \mu_1)^2}{2\sigma^2}},
\end{equation}
where $K_n = K/(mA^2\omega^2)$ is the normalized horizontal kinetic energy per particle. On the other hand, when the alternating transitions occur, the distribution is quite more complex. As demonstrated in Fig. \ref{fig.estadisticaK}(a), these distributions can be well fitted by the addition of two distributions (dashed plus dot-dashed lines in the figure). They show either a bi-modal distribution with two well defined maxima or a bi-modal distribution with two maxima that are quite close but can be still differentiated because one has a large maximum and decreases quickly, and the other has a small maximum and decreases much more slowly. In all cases, the large maximum at low $K$ corresponds to fluctuations of energy close to the minimum value (quasi-absoprbing state) whereas the largest distribution (with or without a clear maximum) corresponds to the fluidized states. We have found a unique distribution that fits well the possibilities just mentioned, namely the addition of a Gaussian (for low $K$) and a  generalized Gaussian. The functional form is 
\begin{eqnarray}
F(K_n) &=& F_1(K_n) + F_2(K_n),\\
F_2(K_n) &=& A_2 e^{-c \sqrt{(K_n - \mu_2)^2 + b^2 }+ d(K_n-\mu_2)},
\end{eqnarray}
where $F_2(K_n)$ can fit a Gaussian for $b \gg (K_n-\mu_2)$ and small $d$ or almost linear exponential decays for $b \ll (K_n - \mu_2)$, which in fact can have different slopes on each side of $\mu_2$.

\begin{figure}[t!]
\includegraphics[width=.95\columnwidth]{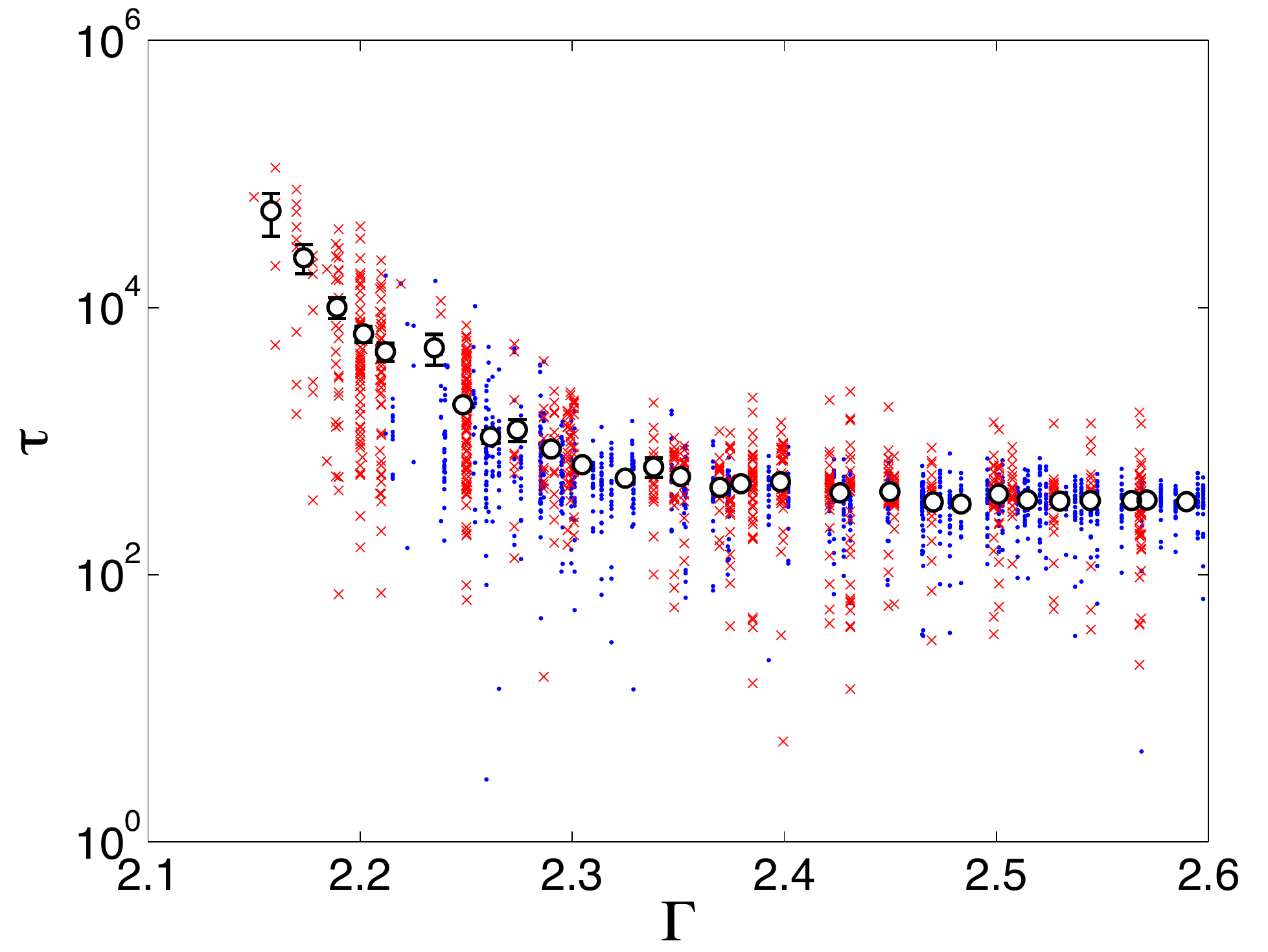}
\caption{(Color online) $\tau$ versus $\Gamma$ obtained from both the horizontal kinetic energy and voltage signals (dots and crosses respectively). Over 2000 fluidization times are shown. Average values are also shown ($\circ$). Errorbars correspond to the standard error definition.}
\label{fig.trtimes.expa}
\end{figure}

We now define the kinetic energy intersection point for each time trace as the particular value $K_{\rm int}$ were the two distributions $F_1$ and $F_2$ intersect, as shown by solid circles in Fig. \ref{fig.estadisticaK}(a). In fact, $K_{\rm int}$ is dependent on $\Gamma$, as shown in in Fig. \ref{fig.estadisticaK}(b). First, the dependence is very strong just near the transition, as both distributions for low and high $K$ tend to be wider, but the dependence is much more weak as $\Gamma$ increases. The weak dependence is phenomenologically fitted to a linear function, $K_{\rm int}/(mA^2\omega^2) = a \Gamma + b$, with $a = -0.079\pm 0.015$ and $b = 0.26 \pm 0.04$. For this fit the two largest $K_{\rm int}$ values were discarded, which are those were the $\Gamma$ dependence is much stronger and with larger variability. We do not have a complete explanation of the observed weak $\Gamma$ dependence. However, it can be understood as the result of two opposite behaviors: as $\Gamma$ increases, the distribution at low $K$ becomes narrower whereas the one at larger $K$ becomes larger and eventually purely exponential, with no clear maximum. Their intersection then occurs at lower $K$ as the acceleration increases. 

From the longer voltage time series the constructed horizontal kinetic energy can also be fitted by the same distributions. Thus, kinetic energy intersection values can also be obtained for computing $\tau$ and its statistics from the voltage signals using the PID control. Figure \ref{fig.trtimes.expa} shows these times obtained from both the horizontal kinetic energy and voltage signals (dots and crosses respectively). Being the voltage signals of longer duration, we find larger $\tau$ values that from the kinetic energy data, but their averages are very close.

\end{document}